\documentclass[aps,nobibnotes,twocolumn,superscriptaddress,showkeys,thelongbibliography]{revtex4-2}
\usepackage{graphicx}

\usepackage{amssymb}
\usepackage{amsmath}
\usepackage{hyperref}
\usepackage{multirow}
\usepackage{placeins}
\usepackage{xcolor}
\usepackage{subcaption}
\newcommand{\bx}{\mathbf{x}}
\newcommand{\bu}{\mathbf{u}}
\newcommand{\pll}{\parallel}

\usepackage{etoolbox}

\begin{document}

\title{Fluid circulation driven by collectively organized metachronal waves in  swimming  {\it T. aceiti} nematodes}

\author{A. C. Quillen}
\email{alice.quillen@rochester.edu}
\affiliation{Department of Physics and Astronomy, University of Rochester, Rochester, NY 14627, USA}
\author{A. Peshkov}
\email{apeshkov@ur.rochester.edu}
\affiliation{Department of Physics and Astronomy, University of Rochester, Rochester, NY 14627, USA}
\author{Brato Chakrabarti}
\email{bchakrabarti@flatironinstitute.org}
\affiliation{Center for Computational Biology, Flatiron Institute,   New York, NY 10010, USA}
\author{Nathan Skerrett}
\email{nskerret@u.rochester.edu}
\affiliation{Department of Physics and Astronomy, University of Rochester, Rochester, NY 14627, USA}
\author{Sonia McGaffigan}
\email{smcgaffi@u.rochester.edu }
\affiliation{Department of Physics and Astronomy, University of Rochester, Rochester, NY 14627, USA}
\author{Rebeca Zapiach}
\email{rzapiach@u.rochester.edu}
\affiliation{Department of Mechanical Engineering, University of Rochester, Rochester, NY 14627, USA}
%
%

\begin{abstract}
Recent experiments have shown that the nematode {\it T. aceti} can assemble into collectively undulating groups at the edge of fluid drops. This coordinated state consists of metachronal waves and drives fluid circulation inside the drop.  We find that the circulation velocity is about 2 mm/s and nearly half the speed of the metachronal wave. We develop a quasi two-dimensional hydrodynamics model using the Stokes flow approximation. The periodic motion of the nematodes constitute our moving boundary condition that drives the flow. Our model suggests that large amplitude excursions of the nematodes tails produce the fluid circulation.    We discuss the constraints on containers that would enhance fluid motion, which could be used in the future design of on demand flow generating systems.

\end{abstract}

\maketitle

\section{Introduction}
\label{sec:intro}

\textit{Turbatrix aceiti} (\textit{T. aceiti}) is a type of freely swimming nematode that have been shown to collectively self-organize at fluid-air interface to form traveling waves \citep{Peshkov_2022,Quillen_2021}.  Individual \textit{T. aceiti} nematodes, also called vinegar eels, are self-propelled swimmers that continuously consume energy. Thus, a dense suspension of vinegar eels is an example of active matter  \citep{Marchetti_2013}. 

Perhaps the most common example of emergent traveling waves is the much studied problem of ciliary carpets. There, hydrodynamic interactions between actively beating cilia, spontaneously result in the formation of large scale waves, known as the metachronal waves \citep{chakrabarti2022multiscale}. Such organized waves are are critical for the motility of ciliated protists (such as \textit{Paramecium} \citep{Tamm_1972}), mucus clearance in mammalian  airways \citep{Sleigh_1988,Afzelius_2004}, and for fluid transport in the brain \citep{faubel2016cilia}. 

What makes our  model system of vinegar eels unique is that unlike cilia which are affixed to a cell membrane, these are freely swimming organisms. They fall under a special class of active agents called `swarmalators' that can self-propel and synchronize their phase of locomotion \citep{OKeeffe_2017}.   It is natural to speculate, can the nematode produced metachronal wave  \citep{Peshkov_2022,Quillen_2021} be harnessed to drive coherent fluid flows?  Key to answering this question are quantitative measurements of the emergent flow, and this is the focus of this paper.

Emergence of coherent fluid pumping states has been reported in both experiments and simulations of other wet active matter systems.  For example, cytoplasmic streaming in plant cells emerges by microfilament self-organization \citep{Woodhouse_2013}.
 With an oil emulsion containing droplets of a highly concentrated aqueous suspension of {\it Bacillus subtilis}, the bacterial suspension organized into a single stable circulating vortex, resulting in fluid pumping \citep{Wioland_2013}.  Collectively formed vortices or mills in plant-animal worms drive fluid flow \citep{Fortune_2021}. 
Simulations of tiny swimming particles (microswimmers) show that schools can corral a volume of liquid much larger than the sum of the volumes swept along by each individual \citep{Jin_2021}.  Examples of active matter driven pumps include design of microfluidic devices that can guide and control motility of self-propelled swimmers resulting in directional flows \citep{Davies_Wykes_2017,Galajda_2007,Guidobaldi_2014}. 

An advantage of studying vinegar eels, compared to many other active systems, is their relatively large size.  Vinegar eels are visible by eye and 1--2 mm in length,  exceeding the size of flagellates, bacteria and many types of cells.  The soil nematode {\it Caenorhabditis elegans}  ({\it C. elegans}) belongs to the same order of {\it Rhabditida} as {\it Turbatrix aceti} and is widely studied as  it is both genomically-defined and amenable to genetic manipulation.  However,  metachronal waves have not been observed in suspensions of {\it C. elegans} \citep{Peshkov_2022}.

In analogy to ciliary transport \cite{Xu_2019}, 
in this paper we seek to understand the relation between the collectively organized traveling waves and the generated fluid flows. We combine experimental observations of dense suspensions of {\it T. aceiti} nematodes with hydrodynamic modeling. The paper is organized as follows: in section \ref{sec:exp} we present our experimental measurements that reveal circulating flow driven by the collective organization of the nematodes in a fluid drop.   In section \ref{sec:hydro} we compute the flow field using a vertical average for Stokes flow and a moving periodic boundary condition.  
In section \ref{subsec:tails} we present experimental observations  of the motions of the tails of nematodes that participate in the metachronal wave. 
The role of the tail motions in influencing the hydrodynamics is explored  in section \ref{subsec:tails_hydro}
and we speculate about the biomechanics of the nematode body motions in section 
\ref{subsec:tail_motions}.
A summary and discussion follows in section \ref{sec:sum}. 

\begin{figure*}[ht] { \centering
\includegraphics[width=2.24 truein,trim=0 -35 0 0,clip]{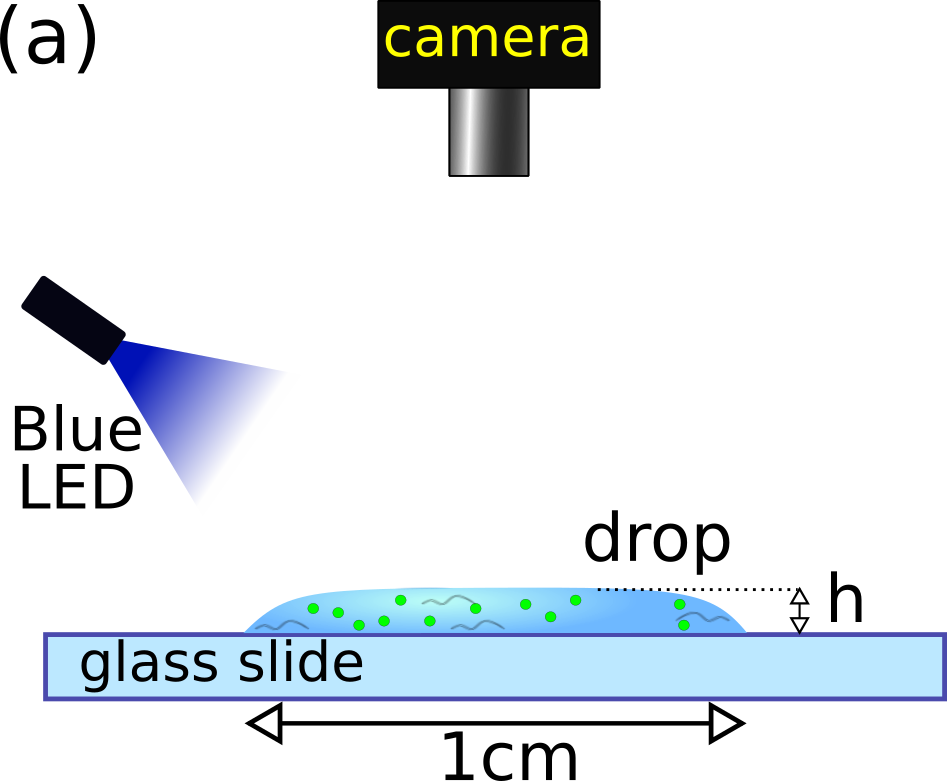}
\includegraphics[width=2.1 truein, trim=15 10 0 0,clip]{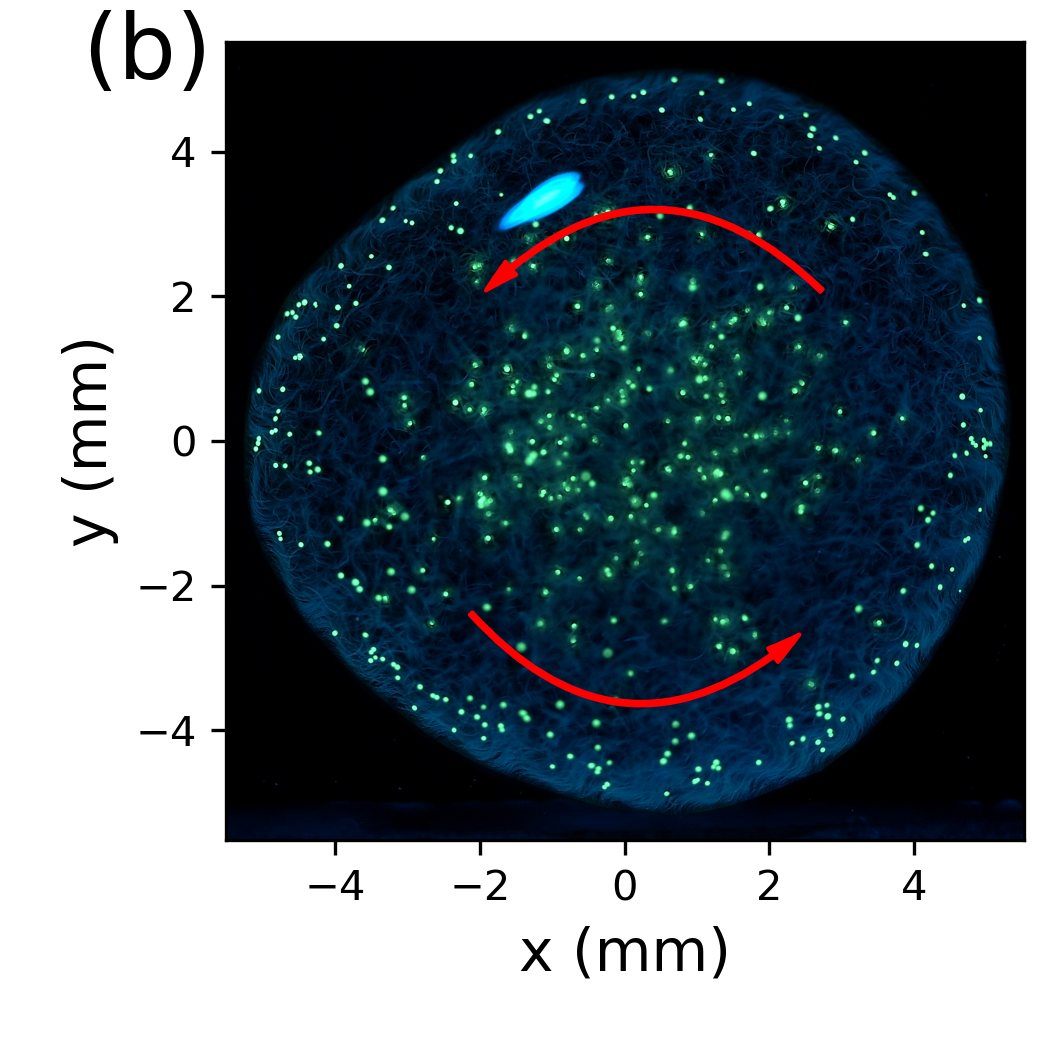}
\includegraphics[width=2.1 truein, trim=15 10 0 0,clip]{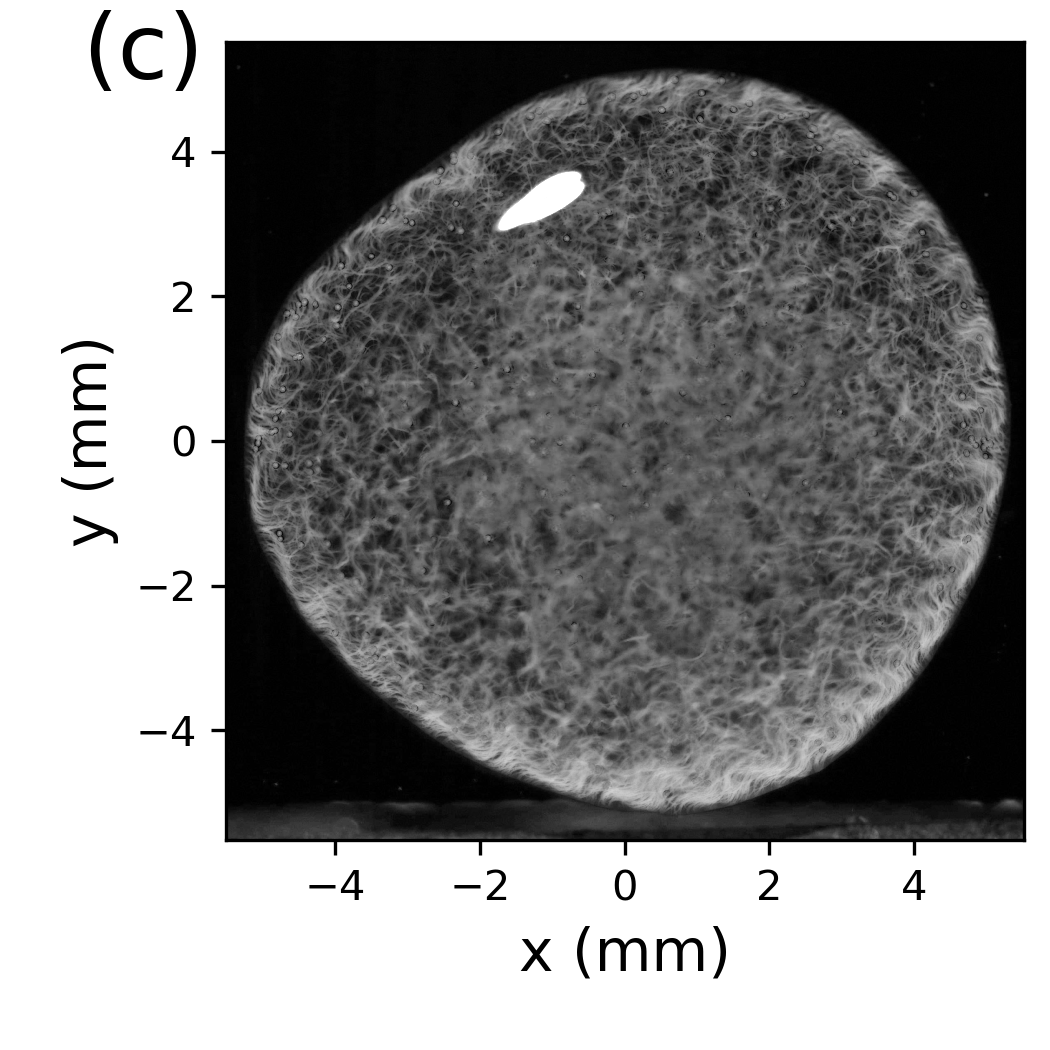}
}
\captionsetup{justification=raggedright, singlelinecheck=false}
\caption{(a) An illustration of the experimental setup. A dilute drop of vinegar contains swimming nematodes and fluorescent microspheres. It is filmed from above with a video camera.  Blue light is absorbed by the microspheres which fluoresce in yellow-green.    The microspheres are used to track circulation induced by collective motion in the vinegar eel population.  (b) A color image from the video which has been annotated to show the direction of circulation.  (c) A postprocessed image from the video shows the nematodes at the boundary.} 
 \label{fig:setup}
\end{figure*}

\section{Experimental methods and Results}
\label{sec:exp}

\subsection{Sample preparation}

We grow the nematodes in a 1:1 solution of distilled water and food grade apple cider vinegar at $\sim$ 5\% concentration. For the experiment, 14 ml of the grow culture containing the nematodes is centrifuged for 3\ min at 5000 rpm. This causes the nematodes to form a dense clump at the bottom of the centrifuge  tube. We extract 300 $\mu l$ of this dense solution and mix it with 10 ml of water and $10$  $\mu l$ of a solution containing fluorescent yellow polyethylene microspheres and a Tween 80 biocompatible surfactant.
The spheres have a diameter of 63--75 $\mu$m and a density of 1.00 g~cm$^{-3}$. The resulting solution is centrifuged again.
Afterwards 100 $\mu l$ of the concentrated
solution was extracted from the bottom of the centrifuge tube and placed onto a bare glass slide. The resulting droplet that we filmed for analysis has a diameter of $\sim 1$ cm and a height $h\approx 1$ mm.

\subsection{Imaging}

The experimental set up for imaging is shown in Fig.~\ref{fig:setup}a
and a video taken with this setup is included as supplemental video A \citep{videoA}.  
The fluorescent microsphere markers are used to measure
the flow induced by the collective motions of the nematodes. We lit the slide with bright blue LEDs, causing the microspheres to fluoresce in yellow-green. 
The drop was filmed in color at 60 frames per second and from above with a Blackmagic Pocket 4K digital camera.

Image frames at a single time from supplemental video A \citep{videoA} are shown in Fig.~\ref{fig:setup}(b) and (c).
In Fig.~\ref{fig:setup}(b),  we show one of the video frames in color.  The microspheres
appear green as they fluorescence and the nematodes appear blue because of the LED lighting.
To best show the nematodes, we subtracted a scaled version of the red frames from the blue ones. 
The resulting image is converted to grayscale and 
shown in Fig.~\ref{fig:setup}(c).  
The nematodes have collected on the boundary, and 
the metachronal wave can be seen in the wave-like features on the outer edge of the drop.  
Higher magnification images (\citep{Quillen_2021}, also discussed below in section \ref{subsec:tails} and shown in supplemental video B \cite{videoB}) 
show that 
the nematode heads are near the boundary and their bodies are oriented 
at an angle with respect to the boundary so that their tails extend into the circulating fluid. 
The bright oval on the top left of the images is a reflection from the lights used
to illuminate the drop and can be ignored.

\begin{figure*}[ht]\centering 
\includegraphics[width=6.2 truein, trim=40 10 10 0,clip]{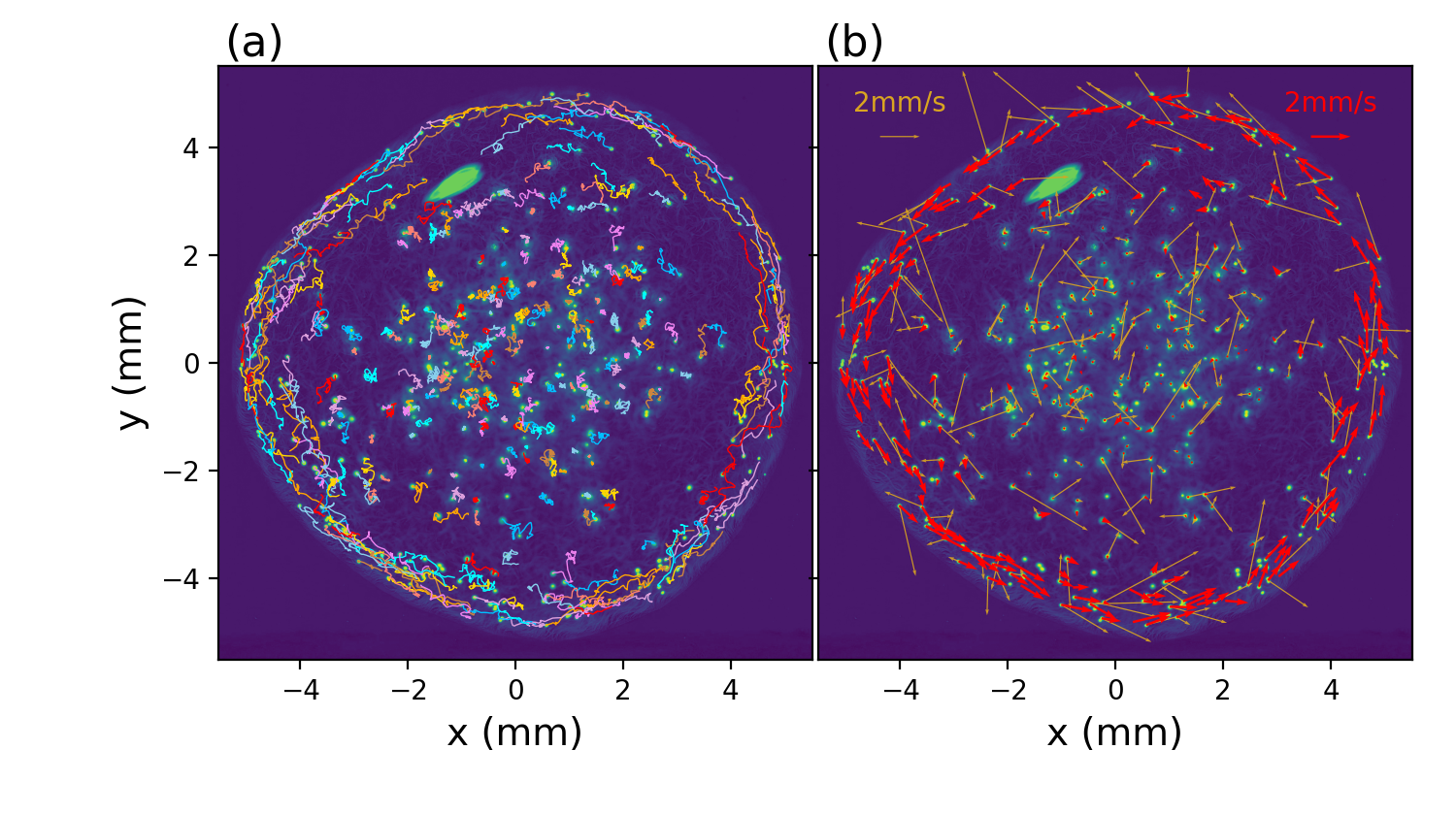}
\captionsetup{justification=raggedright, singlelinecheck=false}
\caption{ 
(a) Tracks of individual fluorescent microspheres from supplemental video A \citep{videoA}.  
The spheres were tracked using 60 video frames during 1 second
of video.  The tracks are shown on top of the first video frame from the video. 
(b)  Velocities of the fluorescent spheres averaged
over 1 second of video are shown with red arrows on top of the first video frame in the sequence.
The thinner brown arrows show instantaneous velocities of the same spheres computed from positions in two consecutive video frames. 
\label{fig:images}}
\end{figure*}

\subsection{Tracking particles}

 The fluorescent microspheres were tracked using the software package \texttt{trackpy} \citep{trackpy},
which implements in Python the Crocker-Grier algorithm for finding and tracking single-particle trajectories \citep{crocker96}.
The microsphere tracers are of sufficiently low concentration that we do not expect them to significantly affect the nematode behavior.

\begin{figure}\centering
\includegraphics[width = 3.4truein, trim = 0 10 -20 0,clip]{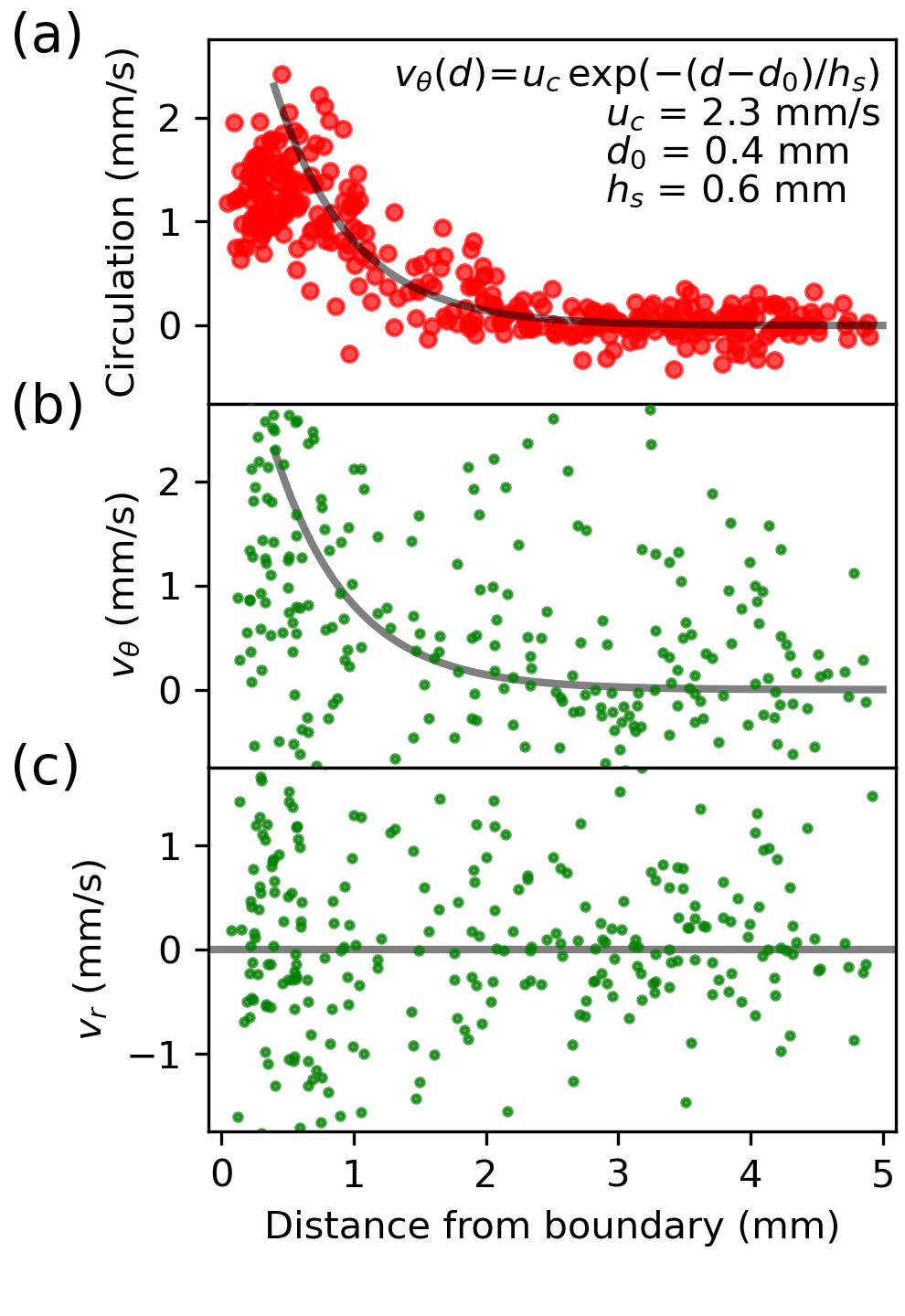}
\captionsetup{justification=raggedright, singlelinecheck=false}
\caption{
Microsphere velocities. 
(a) Average azimuthal velocity component $v_\theta$ of tracked microspheres 
as a function of distance from the outer drop edge.  
The grey line shows an exponentially decaying function with function and parameters
written on the top right of the panel. 
(b) Instantaneous azimuthal velocity component of microspheres as a function of distance from the drop outer edge.
The grey line shows the same exponentially decaying function as in panel (a).
(c) Microsphere instantaneous radial velocity component as a function of distance from the drop outer edge.  The grey line shows a velocity of zero. 
\label{fig:vscatter}
}
\end{figure}

In Fig.~\ref{fig:images}(a) we show
tracks traced by fluorescent microspheres during 1 second of video A \citep{videoA} on top of one of the video frames.
In Fig.~\ref{fig:images}(b) we show velocity vectors computed from these tracks.
In red and with thicker arrows, we show average velocities computed by fitting a line to the trajectory of each microsphere in 1 second of video.
This averages over several oscillations of the metachronal wave to better show fluid 
 circulation.  
Circulation of the microspheres can also be seen directly from viewing supplemental video A  \citep{videoA}.
Instantaneous velocities  are computed from the positions of the particles in the first two frames of  supplemental video A \citep{videoA}.  In Fig.~\ref{fig:images}(b) the instantaneous velocities are shown with thin brown arrows. 

Using a polar coordinate system with origin at the center of the drop, we measure the 
azimuthal and radial velocity components,  $v_\theta, v_r$ for each tracked microsphere. 
In Fig.~\ref{fig:vscatter}(a),  the microsphere azimuthal components of the average velocity (using 1 second of video) are shown as a function of distance $d$ from the outer drop boundary.  
As indicated by the dotted line on the figure, the circulation velocity decays as a function of distance from the boundary and goes to zero at the center of the drop. The black dotted line in on Fig.~\ref{fig:vscatter}(a) shows the curve $v_\theta(d) = u_c e^{-(d-d_0)/h_s}$ with parameters $u_c,d_0 $ and decay length $h_s$.  
The offset $d_0$ is used to describe a peak distance where circulation is highest and $u_c$ gives
the peak circulation velocity.   
Such an exponential decay results from hydrodynamic screening caused flow near boundaries \citep{Liron_1976,Brenner_1999,Cui_2002}.   
In the following section, we will show that the decay length $h_s$ is consisted with hydrodynamic screening in a shallow drop.

In Fig.~\ref{fig:vscatter}(b,c) we plot the instantaneous radial and azimuthal velocity components as a function of distance from the drop edge. 
The exponential curve from Fig.~\ref{fig:vscatter}(a)  is overlayed on Fig.~\ref{fig:vscatter}(b) for comparison and illustrates that the instantaneous velocity can be larger than the time-averaged circulation speed.   
The velocities we measure for the microspheres can exceed the forward swim speed of the nematodes that are involved in the wave.  The nematodes forming the wave advance along the border at a much slower speed, $\sim 0.1$, mm/s than both the metachronal wave and the average azimuthal velocity \citep{Quillen_2021,Peshkov_2022}, highlighting the emergence of large scale coherent transport through collective self-organization. 
The instantaneous velocities are larger than the averaged azimuthal velocity component 
because of oscillations associated with the metachronal wave 
and perturbations caused by close interactions with individual nematodes. 

\subsection{Metachronal wave measurements}

We characterize the kinematics of the emerging metachronal wave. We measure the metachronal wave frequency $f_{MW}$ with a cross correlation technique, as described by \citet{Quillen_2021}.  We compute
the product of two image frames separated by an interval of time and then sum the pixel values in the entire product image.  
The peaks in the sum occur at multiples of the
metachronal wave period.   We similarly measure the metachronal wavelength $\lambda_{MW}$,
by rotating an individual frame about the center of the drop, multiplying it by the original frame and summing over the product image.
Peaks in the sum occur at rotations that are the metachronal wavelength divided by the drop radius.
Errors in these quantities are estimated from the strength and widths of the peaks in these sums.
These measurements are summarized in Table \ref{tab:meas}.
We also list a range of values for the peak averaged azimuthal or tangential speed $u_c$ and 
decay length $h_s$ consistent with the average azimuthal velocities shown Fig.~\ref{fig:vscatter}(a).

\begin{table}
\caption{Measurements \label{tab:meas}}
\begin{tabular}{lll}
\hline
Radius of drop              & $R_d$ & 5.25  mm \\  
Volume of drop & $V_{drop} $ & 100 $\mu$l \\
Metachronal frequency & $f_{MW} $ & $ 5.45 \pm  0.16$ Hz \\ 
Metachronal wavelength & $\lambda_{MW}$ & $0.93 \pm 0.09$ mm \\
Metachronal wave speed & $V_{MW}$ & $ 5.1 \pm 0.5$  mm/s \\
Speed of average flow & $u_{c}$  & $2 \pm 1$ mm/s \\
Exponential decay length of average flow & $h_s$ & $0.7\pm 0.2$ mm \\
\hline
\end{tabular}
\end{table}

The wavelength of the metachronal wave is $\lambda_{MW} \sim 1$ mm, similar
to that of our previous measurements \citep{Quillen_2021}. The metachronal wave
frequency is $f_{MW} \approx 5.45$ Hz, which is within the 4 to 8 Hz range 
measured in similar experiments \citep{Peshkov_2022}. 
Together these give a metachronal wave velocity 
$v_{MW} = \lambda_{MW} f_{MW} \approx 5.1 $ mm/s.
Near the collective wave,  
 the microspheres have a mean averaged azimuthal or tangential velocity component of about $u_c \sim 2$ mm/s.  
The ratio of microsphere mean azimuthal 
velocity component to metachronal wave speed is
 about $u_{c}/v_{MW} \sim 2/5$.

\section{Hydrodynamic theory for coherent flows}\label{sec:hydro}

To understand the underlying mechanics of the coherent transport driven by the metachronal waves we build a quasi 2D hydrodynamic model for the fluid flow.  
 With a flow velocity of 1 mm/s, a length scale of 1 mm and dynamic viscosity of water $\nu \sim 10^{-6} $ m$^2$ s$^{-1}$ = 1 mm$^2$ s$^{-1}$,  the Reynolds number is about 1, so inertia could be important in the velocity field.   Nevertheless, 
we use a model that is appropriate in the Stokes or low Reynolds number limit. 

\subsection{Hele-Shaw approximation for the quasi 2D flow}
\label{sec:depth}

We consider a three-dimensional incompressible flow field in the drop $\mathbf{U} = \{U,V,W\}$. In the limit of low Reynolds number, the evolution of the flow-field is governed by the incompressible Stokes flow equations, 
\begin{align}
	\nabla \cdot \mathbf{U} &=0  \label{eq:inc},  \\ 
	-\frac{1}{\rho}\nabla p + \nu \nabla^2 \mathbf{U} &= \mathbf{0} \label{eq:stok},
\end{align}
where $p$ is the pressure and $\nu$ is the kinematic viscosity of the fluid. The velocity field ${\bf U}$ is a function of $({\bf{x}}_\parallel,z)$, where $\mathbf{x}_\parallel = \{x,y\}$ spans the plane on which the drop resides and the height of the drop $h(\mathbf{x}_\parallel)$ and $0 \leq z \leq h(\bf{x}_\parallel)$.  The characteristic drop thickness $h \sim 1$ mm is  smaller than  the typical drop radius of $R \sim 5$ $\mathrm{mm}$. This natural scale separation allows us to invoke the lubrication approximation  \citep{batchelor2000introduction}. Scaling $\mathbf{x}_\parallel \sim R$ and $z \sim h$ we find $W \sim \{U,V\} h/R$. In the limit $h/R \ll 1$ the vertical velocity $W = 0$ to the leading order and $\partial_z p = 0$ (e.g., \cite{Fortune_2021}). The smallness of the aspect ratio $h/R$ can  be exploited to separate the wall normal dependance (i.e., in the direction $z$) of the velocity field from its in-plane averaged value, as is done in classical Hele-Shaw problems \citep{zeng2003brinkman}. For this purpose, we adopt the following ansatz for the velocity field
\begin{equation}
	\mathbf{U}(\bx_\pll,z) = \bu(\bx_\pll) f(z), \label{eqn:ansatz}
\end{equation}
(following \cite{Fortune_2021})
where $f(z)$ is a single scalar function which describes the velocity profile over the height of the cell and $\bu(\bx_\pll) = u \hat{\mathbf{x}} + v \hat{\mathbf{y}}$ is a 2D depth averaged velocity field defined as
\begin{equation}
	\bu(\bx_\pll) = \frac{1}{h} \int_0^h \mathbf{U} \ \mathrm{d}z.
\end{equation}
The ansatz of Eqn.~\ref{eqn:ansatz} is reasonable because the vertical velocity vanishes to the leading order of the problem.  The form of $f(z)$ depends on the boundary conditions. For flow in the droplet with a free top interface and no-slip bottom surface we take 
\begin{equation}
	    f(z) = \frac{3}{2}\left[1 - \left(1- \frac{z}{h} \right)^2\right]. \label{eqn:fz}
\end{equation}
On the other hand, for flow between two flat plates with no-slip boundary conditions we could use $f(z) = 6 \left(z/h - z^2/h^2\right)$.   The ansatz of Eqn.~\ref{eqn:ansatz} along with the definition
for ${\bu}$ gives 
\begin{equation}
	\partial_{zz} \mathbf{U} = - \alpha_z \frac{\bu}{h^2},
\end{equation}
where $\alpha_z = 3$ for a free interface (consistent with equation \ref{eqn:fz}) and $\alpha_z = 12$ for flow between two no-slip boundaries. In our above formulation we have ignored curvature effects of the drop height near the boundaries. This amounts to assuming that the surface tension effects and the forces from the nematodes in determining the drop shape is negligible (for discussion on these effects see  \cite{Peshkov_2022}). 

We  average Eq.~\eqref{eq:inc} and \eqref{eq:stok} over depth to obtain the quasi 2D approximation for the evolution of the velocity field:
\begin{align}
	 \boldsymbol{\nabla}_\pll \cdot \bu &=0, \\
-\frac{1}{\rho} \boldsymbol{\nabla}_\pll p + \nu \left[\nabla_\pll^2  - \frac{\alpha_z}{h^2} \right] \bu &= \mathbf{0} \label{eqn:NS1},
\end{align}
where $\boldsymbol{\nabla}_\pll = \partial_x \hat{\mathbf{x}} + \partial_y \hat{\mathbf{y}}$ is the 2D gradient and $\nabla_\pll^2$ is the associated 2D Laplacian.
The above set of equations differs from the two-dimensional incompressible Stokes equation by the additional dissipative term $-\alpha_z/h^2$ which accounts for the average friction force between the surface and the droplet interface.  By omitting the term $\nabla_\pll^2 \bu$ we would obtain the so-called Darcy approximation of the Navier-Stokes equation, which is often used to model flows in porous media and viscous fingering in Hele-Shaw cells and describes potential flow. In our case, the flow need not be in general a potential flow \cite{boos1997thermocapillary,bush1997anomalous}. The above set of equations are often termed as the Brinkman correction to Hele-Shaw flows \cite{zeng2003brinkman,Fortune_2021}. It is worth pointing out that, the Green's function associated with the full Stokes equation is non-trivial \citep{Mathijssen_2016} and involves 
the use of the method of images to calculate the flow field \citep{Liron_1976,Mathijssen_2016,Fortune_2021} (see \cite{Fortune_2022} on the validity of approximation).  One advantage of the 2D Brinkman approximation is that it circumvents this challenge and allows us to compute tractable solutions in a spatially periodic domain, as we discuss next.   

Since the depth-averaged velocity $\bu$  acts like a two dimensional incompressible fluid,  we can describe the two-dimensional flow with a stream function $\psi(\bx_\pll)$, where
\begin{equation}
	(u, v) = \left( \partial_y \psi , - \partial_x \psi \right). 
	\label{eqn:uv_stream}
\end{equation}
The vorticity $\omega$ is related to the Laplacian of the stream function 
\begin{equation}
	\omega = \partial_x v- \partial_y u = -  \partial_{xx} \psi -  \partial_{yy} \psi= - \nabla_\pll^2 \psi .
\end{equation}
Upon taking the curl of Eqn.~\ref{eqn:NS1} we obtain the evolution of the vorticity as
\begin{equation}
	\nabla_\pll^2 \omega = \frac{\alpha_z}{h^2} \omega.  \label{eqn:om}
\end{equation}
With negative $\alpha_z$, this is known as  the homogeneous Helmholtz equation, and with positive $\alpha_z$, it is known as the homogeneous screened Poisson equation.  

\begin{figure}[ht]\centering
\includegraphics[width=2truein]{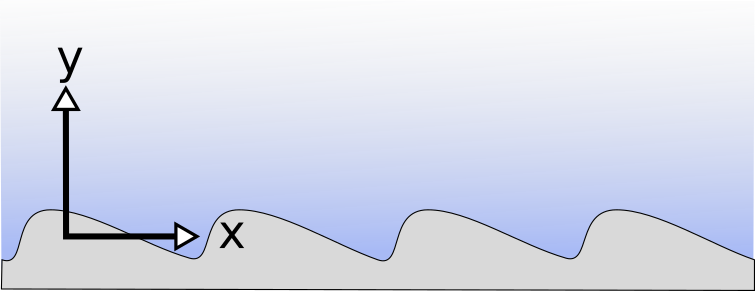}
\captionsetup{justification=raggedright, singlelinecheck=false}
\caption{Coordinate directions.  The boundary is on the bottom
and the direction to the center of the drop is upward. The metachronal wave travels to the right.  We ignore the curvature of the drop boundary. Here the drop is viewed from above. 
\label{fig:flow_xy}}
\end{figure}

As the exponential decay length $h_s$ is smaller than the drop radius $R_d$,  with $h_s/R_d  \approx 0.13 $ we can neglect the curvature of the drop edge. 
We set our $x$ direction along the azimuthal direction of the drop and the $y$ direction is aligned with radius from the drop center and is increasing with distance from the drop edge (see Fig.~\ref{fig:flow_xy}).  Since the flow-field is driven by the motions of the nematode tails we expect the solutions to be periodic along $x$ over a metachronal wavelength. For positive $\alpha_z$,  a general solution of Eqn.~\ref{eqn:om} that is periodic in $x$, with period $\lambda_{MW} = 2 \pi/k_{MW}$,  and decays at larger $y$ has the following stream function 
\begin{align}
\psi(x,y) =&  a_0 e^{- y/h_s}  + \sum_{j=1}^\infty  \Bigg[ \label{eqn:general} \\
&    (a_{cj} \cos (jk_{_{MW}}x) \! + \! a_{sj} \sin (jk_{_{MW}}x) ) e^{-\beta_{j} jk_{_{MW}} y} \nonumber \\
&   (b_{cj} \cos (jk_{_{MW}}x)  \!+ \!b_{sj} \sin (jk_{_{MW}}x) ) e^{- jk_{_{MW}} y} \Bigg] \nonumber,
\end{align}
with coefficients $a_0, a_{cj}, a_{sj}, b_{cj}, b_{sj}$ where 
\begin{equation}
 \beta_{j} \equiv \sqrt{1 + \frac{\alpha_z}{(j k_{_{MW}}h)^2}} .
 \end{equation}
 The characteristic length scale for the exponential  decay associated with the constant circulating term with coefficient $a_0$ is
\begin{equation}
h_s \equiv h/\sqrt{\alpha_z} .\label{eqn:hscreen}
\end{equation}
The terms  with coefficients $b_{cj}, b_{sj}$ have zero vorticity  as they satisfy Laplace's equation $\nabla_\pll^2 \psi = 0$. 

The term proportional to  $\alpha_z$ in Eqn.~\ref{eqn:NS1} gives rise to hydrodynamic screening with an exponential decay length of $h_s = h/\sqrt{\alpha_z}$ typical in the Brinkman correction \cite{brinkman1949calculation}.  With a drop thickness of about $h \sim 1$ mm and $\alpha_z=3$, consistent
with a fixed, no-slip lower boundary and stress free upper interface, the decay length $h_s \sim 0.6 $ mm.
This is consistent with the decay length measured in the induced circulation in section \ref{sec:exp}.

\subsection{The metachronal wave boundary condition} 
\label{sec:boundary}

The nematodes engaged in the metachronal wave are densely packed.  
Their heads are near the outer edge of the boundary and the tails touch moving fluid.   
This suggests that  
 the coherent circulating flow is driven by the motion of the tails of the nematodes involved in the wave.  To incorporate this in our model, we  develop a kinematic description of the tail motion.  We assume that the tails act like a moving, continuous no-slip boundary for the fluid flow. The trajectory of each point on the boundary is described by a periodic function associated with the oscillatory motions in the wave. The metachronal wave gives a delay between the trajectories of neighboring boundary points.

Neglecting curvature of the drop edge, the boundary is spatially periodic in $x$ with wavelength $\lambda_{MW}$
and with metachronal wave traveling in the positive $x$ direction with velocity $V_{MW}$. We use the coordinate $s \in [0,\lambda_{MW})$ to describe positions along the boundary.  The trajectory of a material point at $s=0$ is  described by a displacement function  ${\boldsymbol \delta}_0(t)$ 
which is temporally periodic with frequency $f_{MW}$.  Due to the propagating traveling wave, a point on the boundary that is displaced horizontally from the reference point at $s=0$ undergoes the same trajectory but delayed in time.  With ${\bf x}_b = (x_b,y_b)$,  points on the boundary evolve as 
\begin{align}\label{eqn:b}
{\bf x}_b(s,t) &= {\boldsymbol \delta}_0 (t - s/V_{MW})  + s \hat {\bf x}  + {\bf x}_m .  
\end{align}
In the above description, the boundary is a 1 dimensional space curve that is described by a displacement vector function  ${\boldsymbol \delta}_0(t- s/V_{MW})$. This yields a wave traveling with the metachronal wave velocity $V_{MW}$. The constant ${\bf x}_m$ sets the position of the boundary at $s=t=0$.  The velocity of points on the boundary:
\begin{equation}
{\bf V}_b(s,t) = \frac{d}{dt} \left[ {\boldsymbol \delta}_0 (t - s/V_{MW}) \right].
\end{equation}
The no-slip boundary condition for the fluid velocity  implies:
\begin{align}
{\bf V}_b( s,t)=  {\bf u}({\bf x}_b(s,t)) .  \label{eqn:constraint}
\end{align}

\begin{table}[ht]
\caption{Parameters for flow models \label{tab:flow_models}}
\begin{tabular}{lll}
\hline
Metachronal wavelength & $\lambda_{MW}$ &  1 mm \\
Metachronal frequency  & $f_{MW} $          & 5 Hz \\
Screening parameter   & $\alpha_z$          & 3 \\
Drop thickness             &  $h$                    & 1 mm \\
Screening length   & $h_s = \frac{h}{\sqrt{\alpha_z}}$ & 0.6 mm\\
\hline
\end{tabular}
\end{table}

\subsection{Finding a flow field consistent with the velocity on the boundary}
\label{sec:flow}

Once a set of positions on the boundary and the velocities of these positions are specified, we  try to obtain a two-dimensional fluid flow field consistent with the boundary condition. For this we use a minimization algorithm to find the coefficients in solution of the  stream-function in Eq.~\eqref{eqn:general},  as illustrated below. 
 
Using Eq.~\eqref{eqn:constraint} for the boundary condition and integrating over the boundary, we construct a non-negative function 
\begin{equation}
g({\boldsymbol \delta}_0, \psi) =\frac{1}{L_b} \int_0^{L_b} |{\bf V}_b(s) - {\bf u}(s)|^2 \  \mathrm{d}s \label{eqn:gmin},
\end{equation}
where $L_b$ is the length of the boundary corresponding to a single wavelength of the metachronal wave. The above function is zero for a velocity field $\bu(\bx_\pll)$ derived from a stream function $\psi(\bx_\pll)$ using Eqn.~\ref{eqn:uv_stream}  that is consistent with the velocity boundary condition described by the displacement function ${\boldsymbol \delta}_0(t)$. We use a multivariate minimization algorithm to minimize this function for different values of the  
coefficients in Eq.~\eqref{eqn:general}. For the present problem we retained $j=20$ Fourier modes for the stream-function which results in $81$ free parameters.  To carry out the minimization we used Nelder-Mead Simplex method available through python's \texttt{scipy.optimize} package.  The quality of the result is measured by computing  the standard deviation $\sigma_v$ of the difference between flow velocity and boundary velocity integrated on the boundary.  This is equivalent to the square root of the minimization function,  $\sigma_v = \sqrt{g({\boldsymbol \delta}_0, \psi)} $.  Once the fitting is done, the coefficient $a_0$ of the stream function allows us to compute and characterize the average flow speed  in the model at $y=y_m$ as
\begin{equation}
u_c(y_m) = - \frac{a_0}{h_s} e^{-y_m/h_s}. \label{eqn:uc}
\end{equation}

The key parameters for our boundary model are listed in Table~\ref{tab:flow_models}.  We chose $\alpha_z = 3$ corresponding to a fixed lower boundary and a stress free upper boundary as discussed in section \ref{sec:depth}. We take drop thickness $h=1$ mm which yields a screening length for the solution, $h_s = 0.6$ mm, consistent with
 our experimental measurements.

\begin{figure*}[ht]{\centering
\includegraphics[height=1.15truein,trim = -20 20 -30 0, clip]{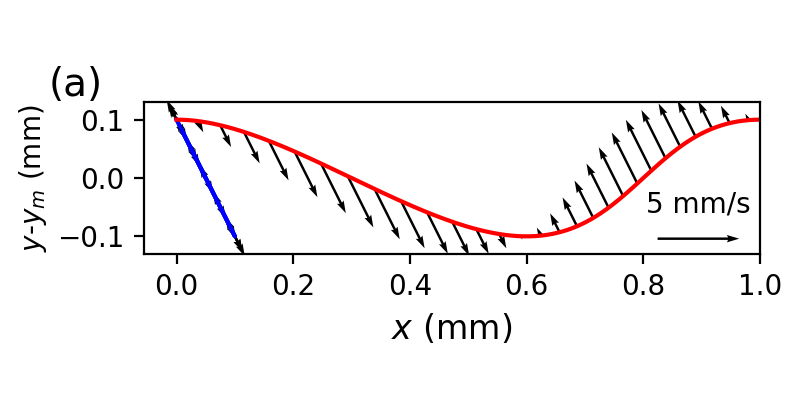}
\includegraphics[height=1.15truein,trim = 17 20 0 0, clip]{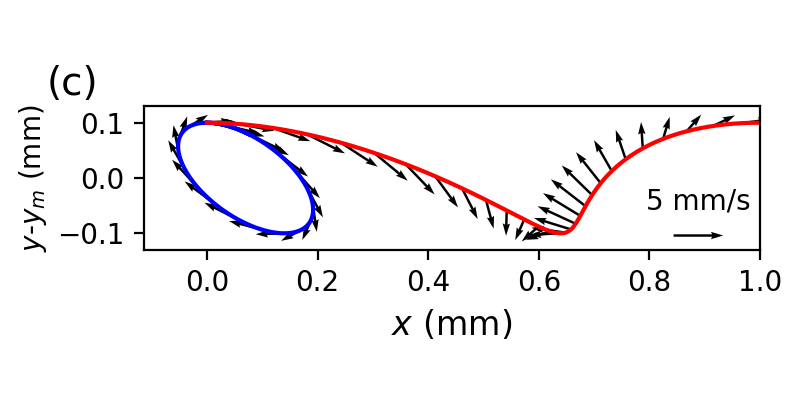} \newline
 \includegraphics[height=3truein,trim = 0 0 0 0, clip]{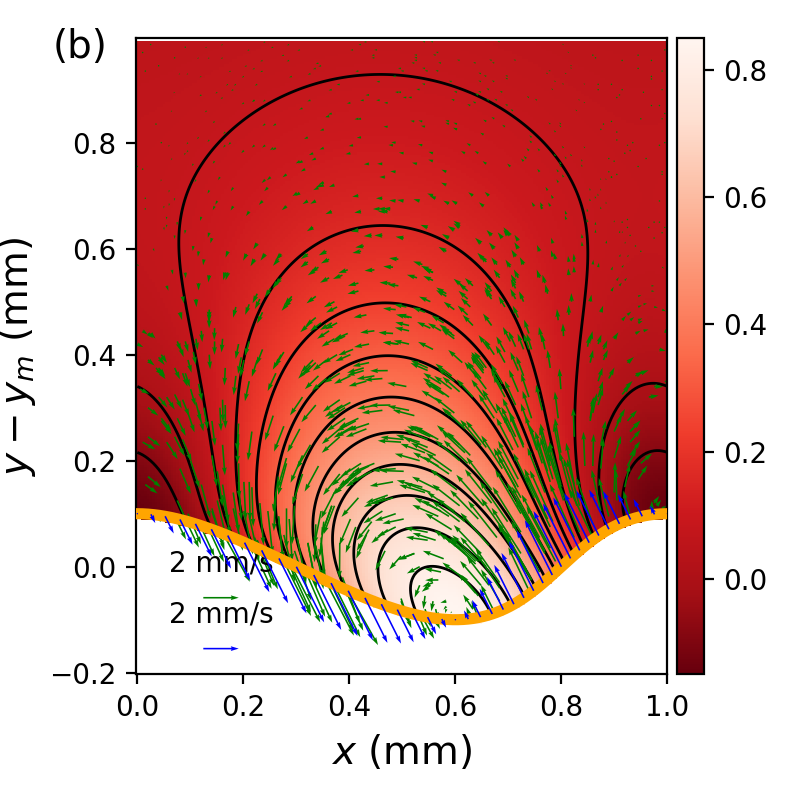}
\includegraphics[height=3truein,trim = 20 0 0 0, clip]{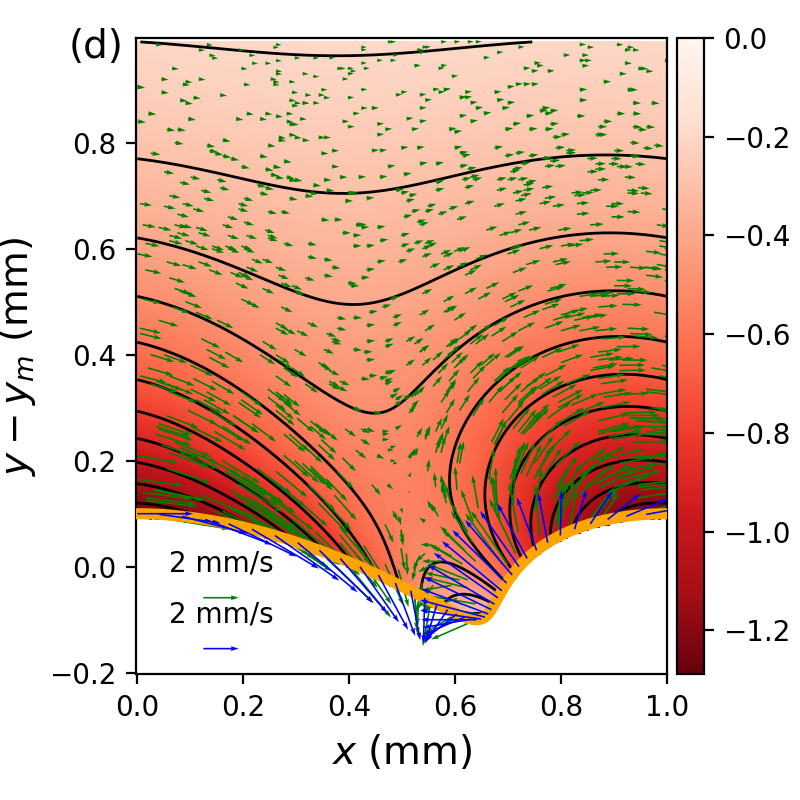}}
\captionsetup{justification=raggedright, singlelinecheck=false}
\caption{(a) In blue we show the position of a single point on the boundary during a full oscillation period
 for a sinusoidal oscillating boundary denoted Model A and with parameters listed in Table~\ref{tab:osc}.
 In red we show the boundary at single moment in time.  Velocity vectors are shown with black arrows. 
(b) A velocity field for Model A that was found by minimizing the function of Eqn.~\ref{eqn:gmin} over the boundary shown 
 in (a).  The stream function is shown as a red image and with black contours.
 Velocity vectors in the fluid are shown with green arrows.   The velocity vectors on
 the boundary are shown with blue arrows.  
(c) Similar to a) but for sinusoidal oscillating boundary model B.   
(d) The associated velocity field for model B. 
 \label{fig:b1b2}
 }
\end{figure*}

\begin{table}[ht]
\caption{Oscillating Boundary Models \label{tab:osc}}
\begin{tabular}{lllll}
\hline
                       &       & Model A   & Model B  & Tail Tr.\\
\hline
Amplitude (mm)      & $A_b$               &     0.10       &    0.10  & -\\
Amplitude  (mm)     &  $B_b$               &     -0.05       &    -0.07  &  - \\ 
Amplitude  (mm)    &   $C_b$               &     0.0       &    0.10    & -\\   
Circulation/flow (mm/s)    & $u_c(y_m)$       &   -0.6        & 1.9  & 2.6 \\
\multicolumn{2}{l}{Visc. dissipation per $\lambda_{MW}$ (pW)} & 70  & 86   & 830 \\
\multicolumn{2}{l}{Power in circulation} &  1\% & 9\%  & 2\% \\
\hline
\end{tabular}
\end{table}

\subsection{Role of different boundary motions on fluid circulation}
\label{sec:oscb}

Motivated by our prior work on a phase oscillator model 
for the collective motion \citep{Quillen_2021}, we consider the velocity field that would arise from a boundary where each point
has a small amplitude of oscillation compared to the metachronal wavelength $\lambda_{MW}$.  We describe the displacement vector of points on the boundary with the function 
\begin{align}\label{eq:bcons}
	{\boldsymbol \delta}_0(t)  =&  A_b \cos( \omega_{MW} t) \hat {\bf y}  +  \\ 
	&  \nonumber\left[B_b  \cos(\omega_{MW} t)   -  C_b  \sin(\omega_{MW} t) \right] \hat {\bf x} \label{eqn:del}
\end{align}
with $\omega_{MW} = 2 \pi f_{MW}$ and three constant coefficients $A_b, B_b, C_b$. We include only three terms because we can adjust the phase of the wave to remove a term that is proportional to $\sin(\omega_{MW} t) \hat {\bf y}$. Positions on the boundary $\bx_b(s,t)$ are then generated from ${\boldsymbol \delta}_0(t)$ using Eq.~\eqref{eqn:b}.  To compute the average flow velocity  (via Eqn.~\ref{eqn:uc}) we take $y_m$ to be the distance between the mean $y$ component of ${\boldsymbol \delta}_0(t)$  and the drop edge.

Depending on the choice of the constants in Eq.~\eqref{eq:bcons}, we obtain qualitatively different models for the boundary motions.   We explore two models, denoted Model A and Model B,  with properties summarized in Table~\ref{tab:osc}.   Model A has $C_b = 0$ and each particle on the boundary moves back and forth along a line segment.  This model is similar to the phase oscillator model \cite{Quillen_2021} developed previously where the points on the nematode bodies moved back and forth, but at an angle with respect to the outer drop edge.  Model B has $C_b \neq 0$ and each particle on the boundary particles has a loop-shaped trajectory.  Model B is inspired by geometric approaches for describing  flows near oscillating boundaries \citep{Purcell_1977,Shapere_1989,Ehlers_2011}.  In this setting a loop trajectory on a surface is described with two infinitesimal operators that don't commute \citep{Purcell_1977},  resulting in a flow velocity that is described via a curvature known as the Stokes curvature
\citep{Shapere_1989,Ehlers_2011}. 
Fig.~\ref{fig:b1b2}(a, c) display the boundary position at a given time instant for the two different boundary models with the arrows indicating the instantaneous velocity of the material points. The blue line on the left shows the trajectory of a single point on the boundary over one oscillation period. 

The associated velocity fields for these two boundary models are depicted in Fig.~\ref{fig:b1b2}(b,d) and their  circulation speeds are listed in Table~\ref{tab:osc}. Interestingly,  Model A, with back and forth oscillation of boundary material points, yields a small circulation speed $u_c$ and predicts the incorrect direction for circulation with $u_c < 0$. In contrast, model B provides 
a circulation speed that is consistent with that we observed experimentally.  The computed solution captures  features of the experimental flow field, including the back and forth motions associated with the metachronal wave.  
Both Model A and B flow fields have rms velocity difference of $\sigma_v = 0.3$ mm/s between flow and boundary (the square root of the minimization function in Eqn.~\ref{eqn:gmin}).

\subsection{Energetics of the flow} 

It is natural to ask how much energy is required to maintain these streaming flows or what fraction of the energy injected by the nematode body motions is used by the coherent flows. To answer this we estimate the viscous dissipation from the velocity fields computed in Fig.~ \ref{fig:b1b2}(b,d) 
(see appendix \ref{sec:power} for details). Using viscosity  $\mu = 0.9 $ mPa~s we find that the estimated dissipation rate is about 80 pW per wavelength for both models of the boundary motion (see Table~\ref{tab:osc}). With about 20 vinegar eels per metachronal wavelength this corresponds to 4 pW per vinegar eel.
Remarkably, we find that for model B, only about 9\% of this power goes into maintaining fluid circulation,  corresponding to the constant or $a_0$ term in the stream function given in Eq.~\eqref{eqn:general} (see Table~\ref{tab:osc}).  The remainder of the power
goes into oscillatory fluid motion. 

\section{Role of tail motions in driving fluid flow}
\label{sec:tails}

The computed flow models for the oscillating boundaries suggest that  elliptical trajectories for points on a boundary are necessary to drive the level of circulation in the fluid that we saw in supplement video A.  As the nematode heads are trapped at the drop edge, the motion is primarily imparted by the nematode tails moving within the circulating fluid. To better understand how the circulation is driven we examine the behavior of tails of the nematodes involved in the metachronal wave.  
\begin{figure*}[ht]\centering
	\includegraphics[height=2.4 truein, trim = 10 0 0 0,clip]{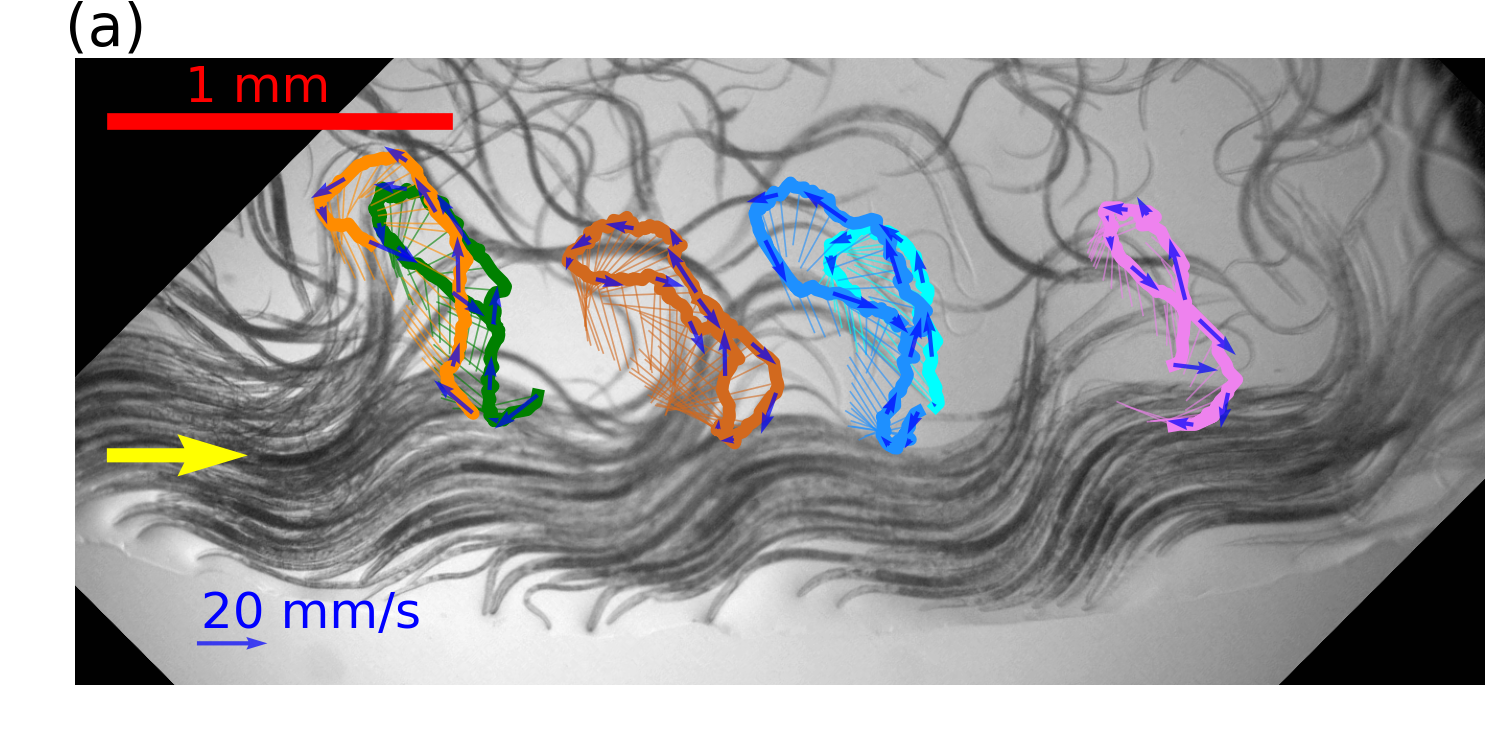}
	\includegraphics[height=2.4 truein,trim = 0 0 0 10,clip]{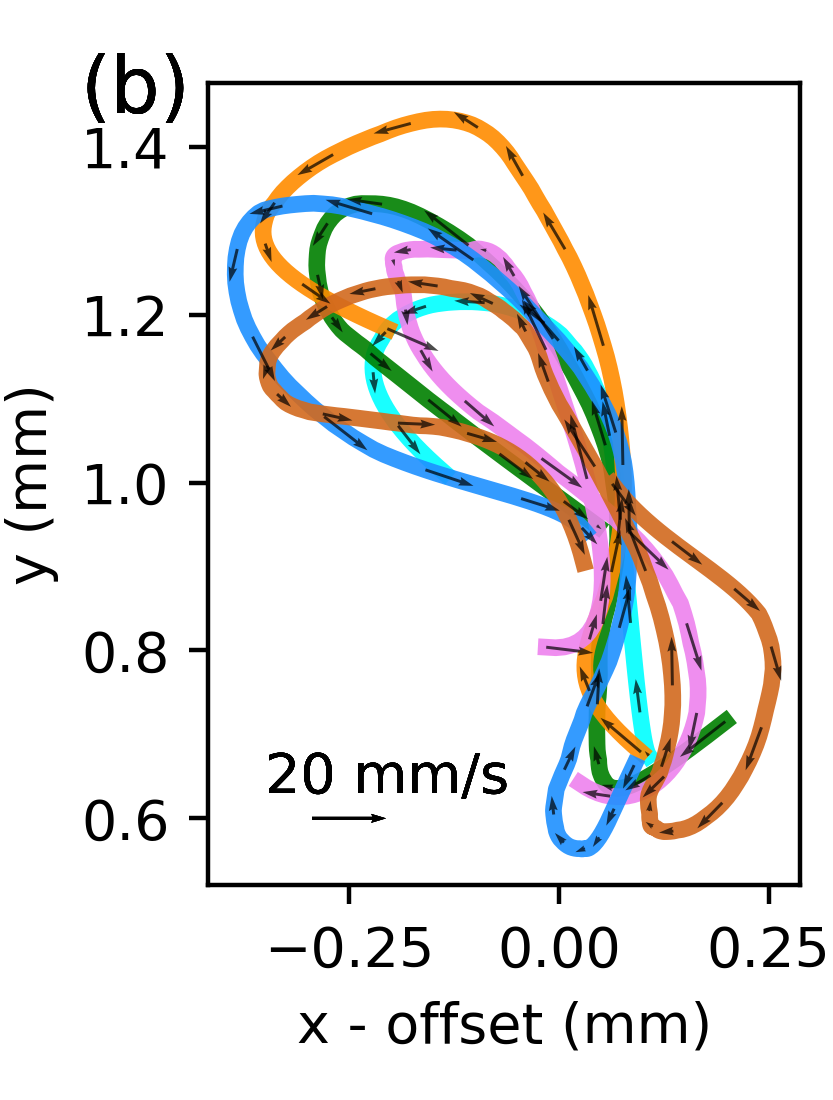}
	\captionsetup{justification=raggedright, singlelinecheck=false}
	\caption{Tail trajectories as seen in a higher magnification high speed video.
		(a) Six nematode tail tips are tracked and plotted with a thick line.  
		Each tail tip is plotted with a different color.   Arrows show the velocity of the tail tip. 
		Thin colored segments show tangent
		directions for the tail at a subset of times with each segment ending at a tail tip location. 
		The underlying gray scale image shows a frame from the high speed video  (supplemental video B; \citep{videoB}). 
		A scale bar is shown in red on the upper left. The yellow arrow shows that  the metachronal wave travels to the right. 
		The heads of vinegar eels involved in the metachronal wave 
		are located at the bottom of the image, near the edge of the drop. 
		(b) Smoothed versions of the same six tracked tail trajectories. 
		Horizontal $x$ positions have been shifted so that the mean position is near zero.  
		Vertical $y$ positions give the distance from the drop edge. 
		\label{fig:tail_tracks} }
\end{figure*}

\subsection{Observed tail trajectories}
\label{subsec:tails}

\begin{figure}[ht]\centering
	\includegraphics[width=2.6 truein, trim =0 12 -20 10, clip]{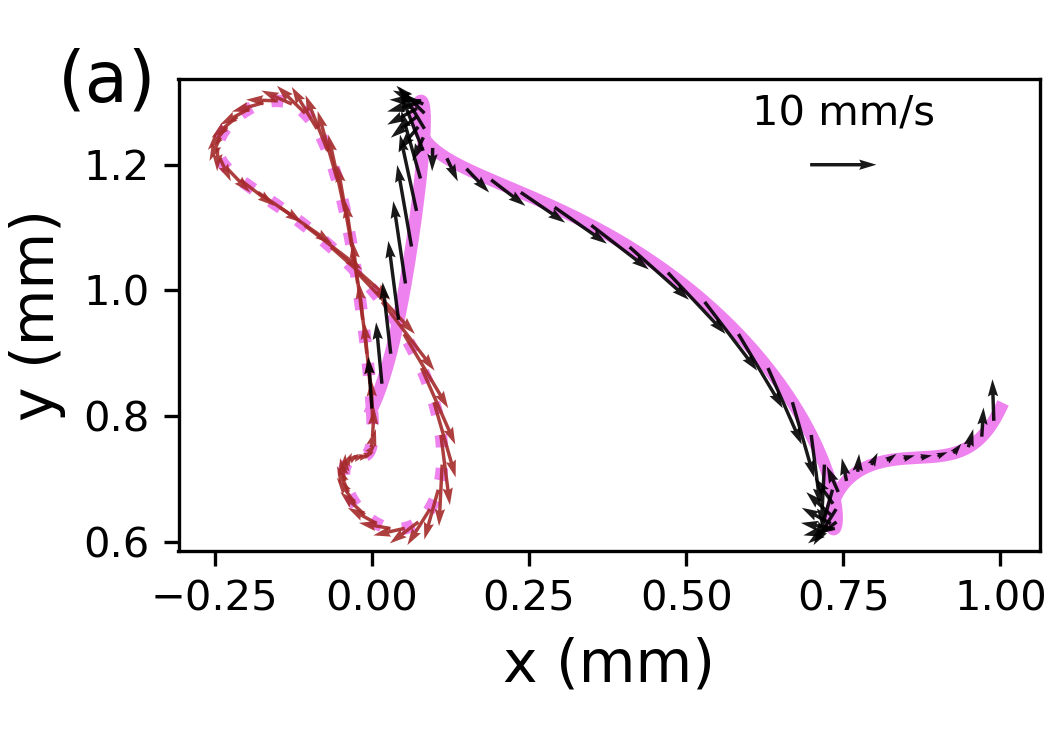}
	\includegraphics[width=3.0 truein, trim = 20 0 20 5, clip]{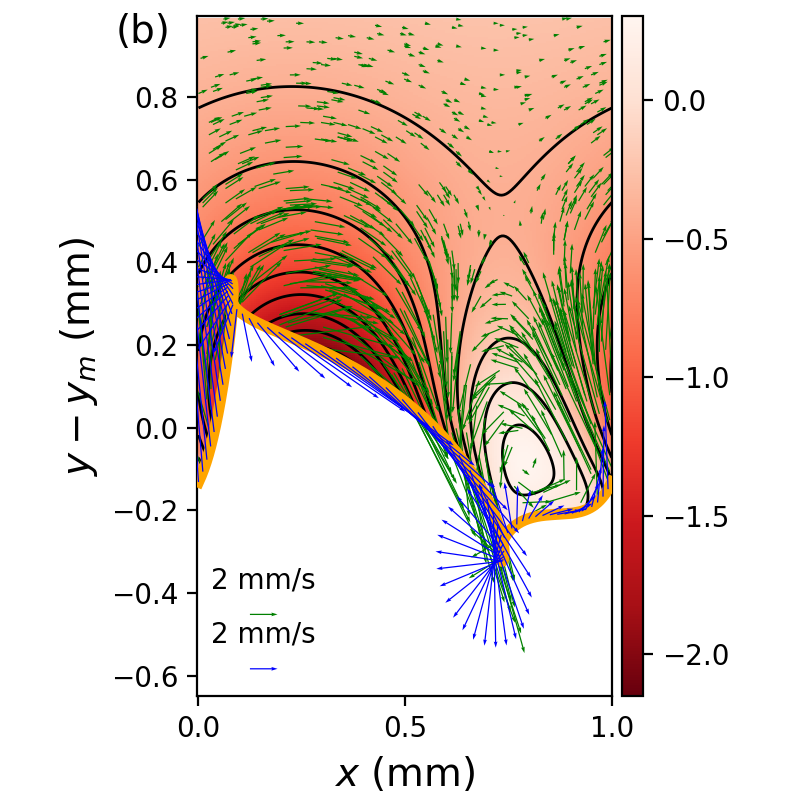}
	\captionsetup{justification=raggedright, singlelinecheck=false}
	\caption{(a)  The trajectory of the rightmost tail in Fig.~ \ref{fig:tail_tracks}
		is shown as a dotted pink line on the left, after smoothing.   Points on the trajectory
		are shifted and delayed using Eqn.~\ref{eqn:b} to give the solid pink line which shows
		the boundary position at a single moment during propagation of the metachronal wave.  
		Arrows show velocity vectors of points on the boundary. 
		(b) The flow field associated with a boundary that is approximated by the tail tip trajectories. 
		The boundary condition is given by points and velocities on the solid line in (a).
		The color map shows the stream function and its contours are plotted with black lines.
		The thick orange line shows the boundary.  Blue vectors show velocities on the boundary.
		Green vectors show fluid velocities. The $y$ axis shows $y-y_m$ where $y_m$ is approximately 
		the mean $y$ position of a boundary point. 
		\label{fig:pink_traj}   }
\end{figure}

To make accurate measurements of the eel tails we use the 10x magnification and high speed videography (1057 fps) described previously  in \citep{Quillen_2021} (see supplemental video B; \citep{videoB}).  
For a few nematode tails, we marked their positions on every frame using the software package \texttt{ImageJ}. The resulting tail trajectories are shown in Fig.~\ref{fig:tail_tracks} and they are also marked on supplemental video B \citep{videoB}.   In Fig.~\ref{fig:tail_tracks}(a), tail tip positions are plotted for about an oscillation period with thick colored lines on top of a gray scale image from supplemental video B.  The tail tip velocities are shown with small arrows. Thin colored segments highlight tangent directions for the tail.   
In Fig.~\ref{fig:tail_tracks}(b) we display smoothed  versions 
of the same six tail tip trajectories.  The trajectories are shifted horizontally so that their mean $x$ positions are the same.  The similarity between the six trajectories gives us confidence that the measured trajectories are associated with the collective motion. 

Based on our previous study on phase oscillator models for the formation of the metachronal wave \citep{Quillen_2021}, we had expected the nematode 
tails to remain near the bodies of other nematodes that are participating in the metachronal wave.   The phase oscillator model assumed that material points on nematode bodies are oscillating back and forth with an amplitude of about 0.1 mm and do not involve large amplitude elliptical motions.  However, as shown in Fig.~\ref{fig:tail_tracks}, and seen in supplemental video B \citep{videoB},  the tracked tail tip trajectories are not ellipses but figure of 8 shapes.   Excursions away from the drop edge (in the $y$ direction on  Fig.~\ref{fig:tail_tracks}) are about 0.8 mm which is almost as large as the metachronal wavelength of 1 mm, and exceeds the back and forth motions of the rest of the nematode bodies and their heads.   The tail tips extend well outside the region near the boundary where the nematodes are densely packed and their velocities can be as high as $\sim 10$ mm/s.

We found that some tails could not be tracked for a whole period because they went out of focus during part of the video.   We note that supplement video B is of a shallow drop, with a depth about half that of the drop shown in supplemental video A; \citep{videoA}.   The shallower depth facilitates viewing individual nematodes, as they are less likely to go in and out of focus,  an issue   at higher magnification.   However, as discussed in section \ref{sec:depth},  a shallower drop depth reduces the screening length $h_s$.  We have no reason to suspect that the depth affects the motions of nematodes participating in the metachronal wave, though the drop shape can influence whether metachronal wave can form \citep{Peshkov_2022}.

\subsection{Flow field computed from tail trajectories}
\label{subsec:tails_hydro}

We proceed to compute the fluid flow resulting from the observed tail trajectories. We use a single tail trajectory to generate a boundary from a sequence of tails undergoing the metachronal wave using Eqn.~\eqref{eqn:b}. On Fig.~\ref{fig:pink_traj}(a) we show the trajectory of the rightmost tail of Fig.~\ref{fig:tail_tracks}(a) with a dotted line.  The associated wave boundary is shown with a solid thick line.  The generated boundary shows that the tail would overlap with neighboring nematodes at different times which is indeed confirmed from the examination of the supplemental video B \citep{videoB} that highlights the large tail tip excursions.  To generate a boundary, 
we chose the rightmost trajectory, shown in pink in Fig.~\ref{fig:tail_tracks}(a),  because it has the least oscillatory motion in the $x$ direction and provides the least overlap.  Reduced overlap in the boundary condition is desirable since it is unphysical to have different fluid velocities at a single point in our 2D fluid model.  Using the boundary shown in Fig.~\ref{fig:pink_traj}(a), we generate an associated two-dimensional fluid flow with the method described in section \ref{sec:flow}. The flow is shown  Fig.~\ref{fig:pink_traj}(b). Circulation (computed from Eq.~\eqref{eqn:uc}) and viscous dissipation rate for this flow model are listed in the rightmost column of Table~\ref{tab:osc}.  The quality of fit, estimated
from Eqn.~\ref{eqn:gmin}, gives rms $\sigma_v \sim 2$ mm/s which is poorer than for 
the flows associated with Model A and B.  The higher value is probably due to the higher  velocities on the boundary and the unphysical overlap region. 

The average flow velocity $u_c  = 2.6$ mm/s, for the flow shown in Fig.~\ref{fig:pink_traj}(b), is sufficiently high to be consistent with the experimental values reported in Fig.~\ref{fig:vscatter}(a).
Integrating horizontally, we find that 
the standard deviation of the $v$ velocity component at $y=0.4$ is 1.6 mm/s.
This is comparable to the standard deviation, 1.1 mm/s, of the instantaneous radial velocity component we measured and showed in Fig.~\ref{fig:vscatter}(c).  Interestingly we find that the power required to drive the flow is about 40 pW per nematode in the wave with only 2\% of this power being used for fluid circulation. This estimated power is about 10 times higher than the 3 pW power estimated power for propulsion of {\it C. elegans}  \citep{Sznitman_2010}, suggesting that vinegar eels expend more energy while they are in the metachronal wave than they would while freely swimming. We speculate that this could be because: (i) tail trajectories for nematodes in the metachronal (extending a maximum distance about 0.8 mm peak to peak) are larger than those of a freely swimming eel (which are about 0.3 mm peak to peak, see the right hand side of Fig.~1 by \citet{Quillen_2021}); (ii) flow velocities may be overestimated because of overlaps in the tail trajectories and (iii) due to confinement, nematodes involved in the metachronal wave would expend part of their energy pushing against each other.   The power required to both drive both the metachronal wave and the associated circulation per nematode could be up to an order of magnitude higher than that expended while freely swimming. Improved hydrodynamic modeling would be needed to better estimate the power required to drive the flow and modeling taking into account steric interactions would be needed to estimate the power dissipated within the nematode bodies. 

\subsection{Mechanical model for a single tail motion}
\label{subsec:tail_motions}

\begin{figure}[ht]
	\centering
	\includegraphics[width=1\linewidth]{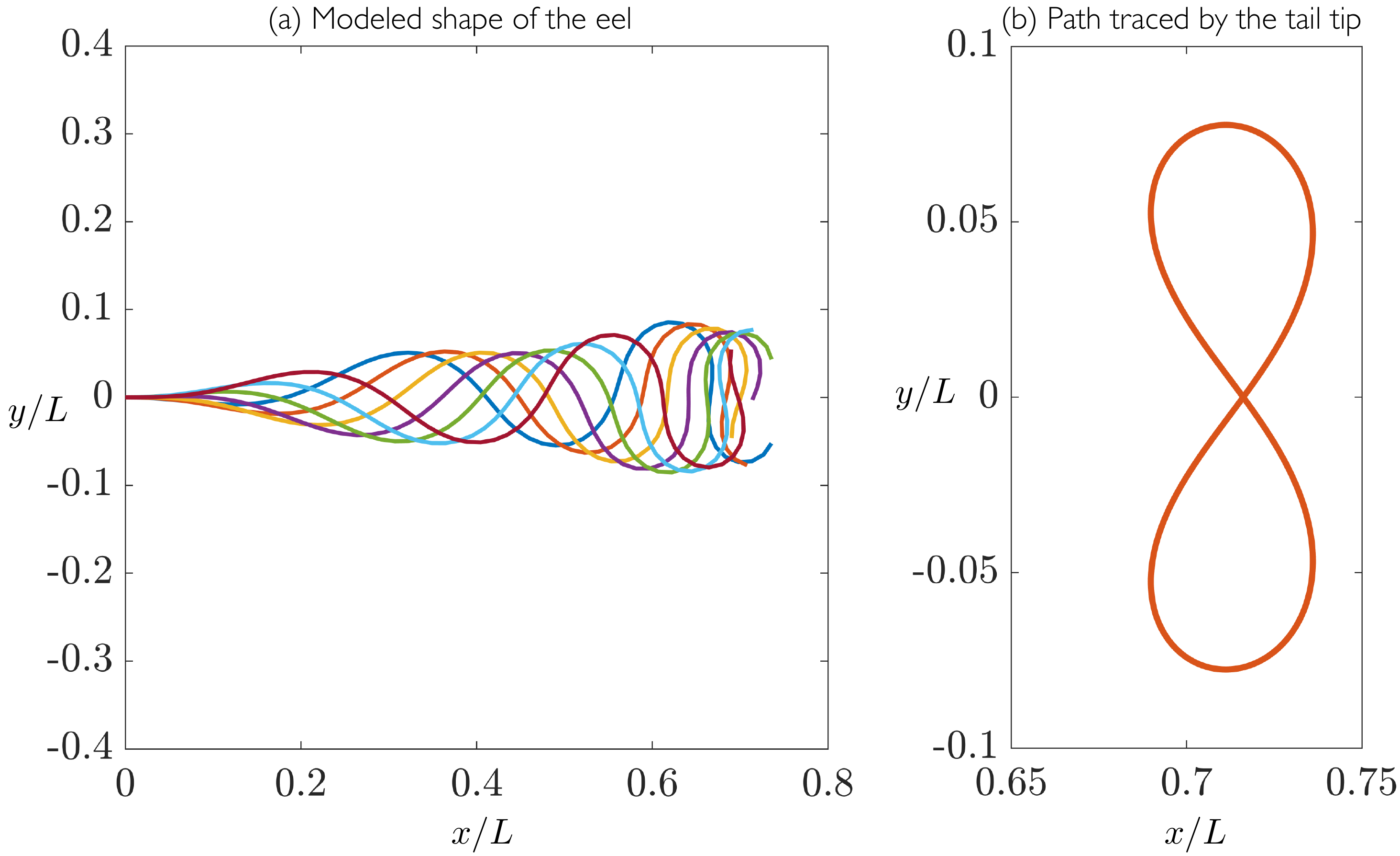}
	\captionsetup{justification=raggedright, singlelinecheck=false}
	\caption{
		(a) Shape of an isolated nematode as computed from the preferred-curvature model. 
		(b) The characteristic figure of 8 shape of the tail tips emerging from the inextensibility of the nematode body.}
	\label{fig:eeltailbrato}
\end{figure}

Our analysis of the nematode tail trajectories in the previous section revealed a characteristic figure of 8 shape for the tail tips. We seek to develop a mechanical model that sheds light on such an emergent pattern. We build a simplified model for the backbone of an isolated nematode. We model the centerline dynamics of an individual nematode as a planar, inextensible Euler elastica with bending rigidity $B$. The centerline is parameterized by arc-length $s$ and identified by a Lagrangian marker $\mathbf{x}(s,t)$.  Following \citep{majmudar2012experiments}, the actuation in the nematode that produces periodic traveling undulation is modeled using a preferred time-dependent curvature given by: 
\begin{equation}
	\kappa_0(s,t) = \frac{A s}{L} \sin \left(k_w s - 2 \pi \Omega_0 t\right).
\end{equation}
Here $k_w$ is the wave number of the target curvature, $A$ is the amplitude and $\Omega_0$ is 
the frequency. We consider the nematode head at $s=0$ to be fixed in space. as the heads of nematodes participating in the metachronal wave do not move very far (see Fig.~5 \cite{Quillen_2021}). 
The tail at $s=L$ is both force and moment free.  Since the vinegar eel is a slender object, we model its hydrodynamics using local slender body theory \cite{KR1976,tornberg2004simulating}, 
which relates the viscous forces $\mathbf{f}_v(s)$ per unit length to the centerline velocity as 
\begin{equation}\label{eq:SBT}
	8\pi\mu \partial_t \mathbf{x}(s,t) = -\mathcal{L}[\mathbf{f}_v].
\end{equation}
Here, $\mu$ is the fluid viscosity, and $\mathcal{L}$ is the local mobility operator that accounts for drag anisotropy along the body and is given by
\begin{equation}
	\mathcal{L}[\mathbf{f}_v](s)=\left[\frac{1}{\xi_\perp}\hat{\mathbf{n}}(s)\hat{\mathbf{n}}(s)+\frac{1}{\xi_\parallel}\hat{\mathbf{t}}(s)\hat{\mathbf{t}}(s)\right]\cdot \mathbf{f}_v(s),
\end{equation}
where $\xi_\perp= (2-c)^{-1}$ and $\xi_\parallel = -(2c)^{-1}$ are resistance coefficients in the normal and tangential directions. For a slender body, coefficient $c$ is assumed to be negative and small. 
In the above expression, $\{\hat{\mathbf{t}}(s),\hat{\mathbf{n}}(s)\}$ are the tangent and normal vectors along the centerline, respectively. The elastic forces are related to the viscous forces following the force and moment balance equations;  
\begin{eqnarray}
	\mathbf{f}_v + \partial_s \mathbf{F} &= \mathbf{0}, \\
	\mathbf{M}_s + \partial_s \mathbf{x} \times \mathbf{F} &= \mathbf{0}.
\end{eqnarray}
Here $\mathbf{F}$ is the elastic contact forces of an inextensible rod, and $\mathbf{M}$ is the bending moment of the filament given by $\mathbf{M}(s) = -B (\kappa(s,t)-\kappa_0(s,t)) \mathbf{x}_s \times \mathbf{x}_{ss}$  (here $\mathbf{x}_s$ refers to the derivative $ \partial_s  \mathbf{x} $). Scaling length by $L$ and time with the relaxation time $\tau = 8 \pi \mu L^4/B$ of an elastic filament, the nematode body shapes are governed by three dimensionless numbers: (i) $A/L$ that sets the amplitude of the target curvature, (ii) $k_w L$ that sets the dimensionless wave number and (iii) $Wr = \Omega_0 \tau$, the worm number \cite{majmudar2012experiments} that compares the frequency of wave propagation to the elastic relaxation time of the nematode body. Using $L = 2$ mm, $B = 10^{-14}$N~m$^2$  \cite{backholm2013viscoelastic}, and $f_0 = 2 \pi \Omega_0 = 5$~Hz we obtain $Wr \approx 40$. We solve the coupled set of PDEs numerically as outlined in \cite{tornberg2004simulating,chakrabarti2019spontaneous}. 

The shapes of the nematode body and the tail trajectories at $\mathbf{x}(s=L)$ are outlined in Fig.~\ref{fig:eeltailbrato} and reveal the natural emergence of the figure of 8 trajectory shape at the tail tip.  A figure of 8 shaped tail tip trajectory can be seen in 
the similar model of an elastic filament (examine Fig.~3b by \cite{Canio_2017}).
Fig.~\ref{fig:eeltailbrato}  also shows that the amplitude of motion reaches a maximum near the tail tip, and this is consistent with observations of both freely swimming nematode and those participating in the metachronal wave.  However, these simulations do not give tail amplitudes as large as we see in either setting.  We compare the largest distance between any two points in the tail trajectory.   This distance is 0.8 mm for the tails tracked in Fig.~\ref{fig:tail_tracks}b, 0.26mm for the freely swimming eel shown in Fig.~1 by \citet{Quillen_2021}, but only 0.16 mm 
in the model tail trajectory shown in Fig.~\ref{fig:eeltailbrato}b.  
The body shapes in Fig.~\ref{fig:eeltailbrato}a  have decreasing wavelength near the tail tip, opposite to what we observe, suggesting that the tails are stiffer or remain straighter than the rest of the nematode bodies.   
When body curvature propagates down a nematode body, as required for locomotion (e.g., \cite{Wen_2012}), we find that a natural outcome is a figure of 8 shaped tail tip trajectory. 
However,  a more complex model than explored here, including steric interactions
and possibly proprioceptive feedback \citep{Wen_2012,Hao_2022}, would be needed to better match the observed body shapes. 
 
\section{Summary and Discussion}
\label{sec:sum}

Using fluorescent microspheres we have measured the circulation velocity of flow in a 1 cm diameter and 
1 mm deep drop that is driven by collective motion in a concentration of swimming {\it T. aceiti} nematodes, which are commonly called vinegar eels. 
The mean circulation velocity is about 2 mm/s,  and located within a mm of the outer edge of the drop.
We find that the ratio of circulating flow velocity to metachronal wave speed is about 2/5 and decays rapidly as a function of distance from the outer drop edge. 

Perturbative models  \citep{Taylor_1951,Brennen_1974,Koiller_1996}
 for causing steady flow by wave-like ripples on a surface have
been developed for three-dimensional flows in the Stokes flow or low Reynolds number limit.
These predict that the driven flow depends on the square of the amplitude of surface perturbations in units of the wavelength 
times the surface wave speed.  This scaling suggests that the ratio of 
flow speed to wave speed should be low.  In this context, the ratio of 2/5 
for the ratio of circulation speed to metachronal wave speed that we measured
in the dilute vinegar drop is unexpectedly high.  

Unlike the planar sheet model \citep{Taylor_1951,Brennen_1974} where the fluid
lies in an infinite 3 dimensional half-space, our
system is shallow, with a fixed lower boundary (the slide that our drop lies on) and free upper boundary (the drop surface).  
We take into account the shallow drop depth by integrating the velocity vertically as a function of depth.  
Depth averaged velocities give a two-dimensional hydrodynamic model.
Neglecting variation of the vertical velocity profiles and drop thickness on horizontal position, 
the fluid obeys the homogeneous screened Poisson equation, typical of Hele-Shaw flow \cite{zeng2003brinkman,Fortune_2021}.

Using the trajectory of a single nematode tail, we construct a boundary condition 
for the metachronal wave by delaying the tail-tip trajectory at neighboring positions. 
This model gives us the velocity and location of particles along a moving 1 dimensional boundary. 
With a general solution to the 2D  homogeneous screened Poisson equation that decays
at large distances from the boundary and using an optimization routine,
we find the stream function that minimizes the difference between fluid velocity
and particle velocities on the boundary. 
The screening in the screened Poisson equation is sensitive to the drop depth,
with shallower drops having circulating flows that decay more rapidly as a function
of distance from the moving boundary than flows in deeper drops. 
We find that 
the circulation or streaming velocity is sensitive to the boundary particle trajectory shape, with ellipse trajectories
more effective than linear trajectories at driving circulation. 

We had expected to find that the tails of nematodes participating the metachronal wave undergo ellipse trajectories.
However, upon examination, we found that the tails of the nematodes undergoing collective motion have figure of 8 trajectories and undergo large 0.8 mm excursions away from the drop edge. These are about twice as large as tail motions exhibited by a freely swimming nematode or most of the nematode body that is participating in the metachronal wave. 

To probe the mechanical origins of the tail motions we constructed a simple preferred curvature model for an isolated nematode. 
This model captures the characteristic figure of 8 shape of the tail and shows that this shape is naturally caused by a curvature wave propagating down an elastic body.   While this model captures the characteristic figure of 8 shape of the tail and an increase in amplitude toward the tail tip, the simulated results are not in great agreement for an isolated nematode. This can be possibly due to subtle structural features associated within the nematode such as varying cross-sectional width along its backbone and variations in how such structural features alter the internal force-generation. The associated biomechanics will be considered in the future. Understanding collective effects poses another set of challenges in modeling. For example, in contrast with the experimental measurements for nematodes
participating in the metachronal wave, our simple mechanical model does not predict large amplitude oscillations of the tail tip. Since hydrodynamic interactions are screened, we believe that such emergent dynamics of the nematode tails could be associated with local steric interactions.
Our prior work suggested that steric interactions were required for metachronal wave formation in the vinegar eel system and caused the waves traveling down the nematode bodies to deviate from a pure sinusoidal function \citep{Quillen_2021}. 
Recently, it has been proposed that steric interactions alone can lead to the emergence of collective behaviors in arrays of cilia and alter their isolated waveforms \cite{chelakkot2021synchronized}.  The potential role of such effects remain to be explored. 

A flow model generated with a boundary generated from the observed tail motions exhibits sufficient circulation to be consistent with the circulation we observed experimentally.  The flow model suggests that the motions of the vinegar eel tails are important for driving fluid circulation. 

The high ratio of circulation flow speed to metachronal wave speed that we measured  suggests that systems of nematodes could be engineered to drive flows. 
What type of container would best give fluid circulation?
Bordertaxis of the nematodes should be facilitated, otherwise the nematodes
will not enter a collective state giving a metachronal wave.  The container edge where the eels would be corralled could be shallow (less than 1 mm thick) or beveled, 
similar to the contact angle
caused by surface tension in a drop (with contact angle below 70$^\circ$; see \citet{Peshkov_2022}). 
The top surface could be covered, rather than open, though this would affect 
the screening length scale.  We assumed an open surface giving $\alpha_z \sim 3$
and a screening exponential length $h_s = h/\sqrt{\alpha_z}$ (as  
in Eqn.~\ref{eqn:hscreen}) where $h$ is the depth of the container.  
With a fixed upper boundary (for an enclosed container) we would expect $\alpha_z \sim 12$.
A container twice as deep would
be needed to give the same screening length in a closed container as one with an open surface. 
A beveled edge might be the best compromise as it might corral the eels, facilitating
collective motion, while simultaneously allowing a larger screening exponential decay length for the driven  flow.   
Large amplitude motions in the nematode tails would help  drive circulation, so the distance between outer and inner container edges should be at least a few mm. 

 
We observed nematode bodies overlapping other nematodes in our high magnification
video and there are numerous nematodes that are not engaged in the collective motion.  Our 2D flow and boundary model neglects
motions of overlapping nematodes and those that do not participate in the collective motion.  Extending the fidelity and capability of the hydrodynamic modeling is challenging but would improve understanding of the relation between collective motion in a dense concentrations of oscillating swimming organisms and associated flows resulting from their collective motion.

\

\vskip 1 truein

\
 
\begin{acknowledgments}

This material is based upon work supported in part  by 
National Science Foundation Grant No. PHY-1757062, and
National Science Foundation Grant No. DMR-1809318.

We thank Michael J. Shelley and Adam Lamson for helpful discussions and suggestions.  We thank the Flatiron Institute, Center for Computational Biology for their welcome and hospitality in Spring 2022. 
We thank Maya Parada for many interesting discussions on the subject of designing and manufacturing containers for the vinegar eels. 

\end{acknowledgments}





\bibliography{eels}

\begin{thebibliography}{47}
\expandafter\ifx\csname natexlab\endcsname\relax\def\natexlab#1{#1}\fi
\expandafter\ifx\csname bibnamefont\endcsname\relax
  \def\bibnamefont#1{#1}\fi
\expandafter\ifx\csname bibfnamefont\endcsname\relax
  \def\bibfnamefont#1{#1}\fi
\expandafter\ifx\csname citenamefont\endcsname\relax
  \def\citenamefont#1{#1}\fi
\expandafter\ifx\csname url\endcsname\relax
  \def\url#1{\texttt{#1}}\fi
\expandafter\ifx\csname urlprefix\endcsname\relax\def\urlprefix{URL }\fi
\providecommand{\bibinfo}[2]{#2}
\providecommand{\eprint}[2][]{\url{#2}}

\bibitem[{\citenamefont{Peshkov et~al.}(2022)\citenamefont{Peshkov, McGaffigan,
  and Quillen}}]{Peshkov_2022}
\bibinfo{author}{\bibfnamefont{A.}~\bibnamefont{Peshkov}},
  \bibinfo{author}{\bibfnamefont{S.}~\bibnamefont{McGaffigan}},
  \bibnamefont{and} \bibinfo{author}{\bibfnamefont{A.~C.}
  \bibnamefont{Quillen}}, \bibinfo{journal}{Soft Matter}
  \textbf{\bibinfo{volume}{18}}, \bibinfo{pages}{1174} (\bibinfo{year}{2022}).

\bibitem[{\citenamefont{Quillen et~al.}(2021)\citenamefont{Quillen, Peshkov,
  Wright, and McGaffigan}}]{Quillen_2021}
\bibinfo{author}{\bibfnamefont{A.~C.} \bibnamefont{Quillen}},
  \bibinfo{author}{\bibfnamefont{A.}~\bibnamefont{Peshkov}},
  \bibinfo{author}{\bibfnamefont{E.}~\bibnamefont{Wright}}, \bibnamefont{and}
  \bibinfo{author}{\bibfnamefont{S.}~\bibnamefont{McGaffigan}},
  \bibinfo{journal}{Phys. Rev. E} \textbf{\bibinfo{volume}{104}},
  \bibinfo{pages}{014412} (\bibinfo{year}{2021}).

\bibitem[{\citenamefont{{Marchetti} et~al.}(2013)\citenamefont{{Marchetti},
  {Joanny}, {Ramaswamy}, {Liverpool}, {Prost}, {Rao}, and
  {Simha}}}]{Marchetti_2013}
\bibinfo{author}{\bibfnamefont{M.~C.} \bibnamefont{{Marchetti}}},
  \bibinfo{author}{\bibfnamefont{J.~F.} \bibnamefont{{Joanny}}},
  \bibinfo{author}{\bibfnamefont{S.}~\bibnamefont{{Ramaswamy}}},
  \bibinfo{author}{\bibfnamefont{T.~B.} \bibnamefont{{Liverpool}}},
  \bibinfo{author}{\bibfnamefont{J.}~\bibnamefont{{Prost}}},
  \bibinfo{author}{\bibfnamefont{M.}~\bibnamefont{{Rao}}}, \bibnamefont{and}
  \bibinfo{author}{\bibfnamefont{R.~A.} \bibnamefont{{Simha}}},
  \bibinfo{journal}{Reviews of Modern Physics} \textbf{\bibinfo{volume}{85}},
  \bibinfo{pages}{1143} (\bibinfo{year}{2013}).

\bibitem[{\citenamefont{Chakrabarti et~al.}(2022)\citenamefont{Chakrabarti,
  Furthauer, and Shelley}}]{chakrabarti2022multiscale}
\bibinfo{author}{\bibfnamefont{B.}~\bibnamefont{Chakrabarti}},
  \bibinfo{author}{\bibfnamefont{S.}~\bibnamefont{Furthauer}},
  \bibnamefont{and} \bibinfo{author}{\bibfnamefont{M.~J.}
  \bibnamefont{Shelley}}, \bibinfo{journal}{Proceedings of the National Academy
  of Sciences} \textbf{\bibinfo{volume}{119}}, \bibinfo{pages}{e2113539119}
  (\bibinfo{year}{2022}).

\bibitem[{\citenamefont{Tamm}(1972)}]{Tamm_1972}
\bibinfo{author}{\bibfnamefont{S.~L.} \bibnamefont{Tamm}},
  \bibinfo{journal}{The Journal of Cell Biology} \textbf{\bibinfo{volume}{55}},
  \bibinfo{pages}{250} (\bibinfo{year}{1972}).

\bibitem[{\citenamefont{Sleigh et~al.}(1988)\citenamefont{Sleigh, Blake, and
  Liron}}]{Sleigh_1988}
\bibinfo{author}{\bibfnamefont{M.~A.} \bibnamefont{Sleigh}},
  \bibinfo{author}{\bibfnamefont{J.~R.} \bibnamefont{Blake}}, \bibnamefont{and}
  \bibinfo{author}{\bibfnamefont{N.}~\bibnamefont{Liron}},
  \bibinfo{journal}{American Review of Respiratory Disease}
  \textbf{\bibinfo{volume}{137}}, \bibinfo{pages}{726} (\bibinfo{year}{1988}).

\bibitem[{\citenamefont{Afzelius}(2004)}]{Afzelius_2004}
\bibinfo{author}{\bibfnamefont{B.~A.} \bibnamefont{Afzelius}},
  \bibinfo{journal}{Journal of Pathology} \textbf{\bibinfo{volume}{204}},
  \bibinfo{pages}{470} (\bibinfo{year}{2004}).

\bibitem[{\citenamefont{Faubel et~al.}(2016)\citenamefont{Faubel, Westendorf,
  Bodenschatz, and Eichele}}]{faubel2016cilia}
\bibinfo{author}{\bibfnamefont{R.}~\bibnamefont{Faubel}},
  \bibinfo{author}{\bibfnamefont{C.}~\bibnamefont{Westendorf}},
  \bibinfo{author}{\bibfnamefont{E.}~\bibnamefont{Bodenschatz}},
  \bibnamefont{and} \bibinfo{author}{\bibfnamefont{G.}~\bibnamefont{Eichele}},
  \bibinfo{journal}{Science} \textbf{\bibinfo{volume}{353}},
  \bibinfo{pages}{176} (\bibinfo{year}{2016}).

\bibitem[{\citenamefont{O'Keeffe et~al.}(2017)\citenamefont{O'Keeffe, Hong, and
  Strogatz}}]{OKeeffe_2017}
\bibinfo{author}{\bibfnamefont{K.~P.} \bibnamefont{O'Keeffe}},
  \bibinfo{author}{\bibfnamefont{H.}~\bibnamefont{Hong}}, \bibnamefont{and}
  \bibinfo{author}{\bibfnamefont{S.~H.} \bibnamefont{Strogatz}},
  \bibinfo{journal}{Nature Communications} \textbf{\bibinfo{volume}{8}},
  \bibinfo{pages}{1504} (\bibinfo{year}{2017}),
  \urlprefix\url{https://doi.org/10.1038%2Fs41467-017-01190-3}.

\bibitem[{\citenamefont{Woodhouse and Goldstein}(2013)}]{Woodhouse_2013}
\bibinfo{author}{\bibfnamefont{F.~G.} \bibnamefont{Woodhouse}}
  \bibnamefont{and} \bibinfo{author}{\bibfnamefont{R.~E.}
  \bibnamefont{Goldstein}}, \bibinfo{journal}{PNAS}
  \textbf{\bibinfo{volume}{110}}, \bibinfo{pages}{14132}
  (\bibinfo{year}{2013}).

\bibitem[{\citenamefont{Wioland et~al.}(2013)\citenamefont{Wioland, Woodhouse,
  Dunkel, Kessler, and Goldstein}}]{Wioland_2013}
\bibinfo{author}{\bibfnamefont{H.}~\bibnamefont{Wioland}},
  \bibinfo{author}{\bibfnamefont{F.~G.} \bibnamefont{Woodhouse}},
  \bibinfo{author}{\bibfnamefont{J.}~\bibnamefont{Dunkel}},
  \bibinfo{author}{\bibfnamefont{J.~O.} \bibnamefont{Kessler}},
  \bibnamefont{and} \bibinfo{author}{\bibfnamefont{R.~E.}
  \bibnamefont{Goldstein}}, \bibinfo{journal}{Phys. Rev. Lett.}
  \textbf{\bibinfo{volume}{110}}, \bibinfo{pages}{268102}
  (\bibinfo{year}{2013}).

\bibitem[{\citenamefont{Fortune et~al.}(2021)\citenamefont{Fortune, Worley,
  Sendova-Franks, Franks, Leptos, Lauga, and Goldstein}}]{Fortune_2021}
\bibinfo{author}{\bibfnamefont{G.~T.} \bibnamefont{Fortune}},
  \bibinfo{author}{\bibfnamefont{A.}~\bibnamefont{Worley}},
  \bibinfo{author}{\bibfnamefont{A.~B.} \bibnamefont{Sendova-Franks}},
  \bibinfo{author}{\bibfnamefont{N.~R.} \bibnamefont{Franks}},
  \bibinfo{author}{\bibfnamefont{K.~C.} \bibnamefont{Leptos}},
  \bibinfo{author}{\bibfnamefont{E.}~\bibnamefont{Lauga}}, \bibnamefont{and}
  \bibinfo{author}{\bibfnamefont{R.~E.} \bibnamefont{Goldstein}},
  \bibinfo{journal}{J. Fluid Mech.} \textbf{\bibinfo{volume}{914}},
  \bibinfo{pages}{A20} (\bibinfo{year}{2021}).

\bibitem[{\citenamefont{Jin et~al.}(2021)\citenamefont{Jin, Chen, Maass, and
  Mathijssen}}]{Jin_2021}
\bibinfo{author}{\bibfnamefont{C.}~\bibnamefont{Jin}},
  \bibinfo{author}{\bibfnamefont{Y.}~\bibnamefont{Chen}},
  \bibinfo{author}{\bibfnamefont{C.~C.} \bibnamefont{Maass}}, \bibnamefont{and}
  \bibinfo{author}{\bibfnamefont{A.~J.} \bibnamefont{Mathijssen}},
  \bibinfo{journal}{Phys. Rev. Lett.} \textbf{\bibinfo{volume}{127}},
  \bibinfo{pages}{088006} (\bibinfo{year}{2021}).

\bibitem[{\citenamefont{Davies~Wykes et~al.}(2017)\citenamefont{Davies~Wykes,
  Zhong, Tong, Adachi, Liu, Ristroph, Ward, Shelley, and
  Zhang}}]{Davies_Wykes_2017}
\bibinfo{author}{\bibfnamefont{M.~S.} \bibnamefont{Davies~Wykes}},
  \bibinfo{author}{\bibfnamefont{X.}~\bibnamefont{Zhong}},
  \bibinfo{author}{\bibfnamefont{J.}~\bibnamefont{Tong}},
  \bibinfo{author}{\bibfnamefont{T.}~\bibnamefont{Adachi}},
  \bibinfo{author}{\bibfnamefont{Y.}~\bibnamefont{Liu}},
  \bibinfo{author}{\bibfnamefont{L.}~\bibnamefont{Ristroph}},
  \bibinfo{author}{\bibfnamefont{M.~D.} \bibnamefont{Ward}},
  \bibinfo{author}{\bibfnamefont{M.~J.} \bibnamefont{Shelley}},
  \bibnamefont{and} \bibinfo{author}{\bibfnamefont{J.}~\bibnamefont{Zhang}},
  \bibinfo{journal}{Soft Matter} \textbf{\bibinfo{volume}{13}},
  \bibinfo{pages}{4681} (\bibinfo{year}{2017}).

\bibitem[{\citenamefont{Galajda et~al.}(2007)\citenamefont{Galajda, Keymer,
  Chaikin, and Austin}}]{Galajda_2007}
\bibinfo{author}{\bibfnamefont{P.}~\bibnamefont{Galajda}},
  \bibinfo{author}{\bibfnamefont{J.}~\bibnamefont{Keymer}},
  \bibinfo{author}{\bibfnamefont{P.}~\bibnamefont{Chaikin}}, \bibnamefont{and}
  \bibinfo{author}{\bibfnamefont{R.}~\bibnamefont{Austin}},
  \bibinfo{journal}{J. Bacteriology} \textbf{\bibinfo{volume}{189}},
  \bibinfo{pages}{8704} (\bibinfo{year}{2007}).

\bibitem[{\citenamefont{Guidobaldi et~al.}(2014)\citenamefont{Guidobaldi,
  Jeyaram, Berdakin, Moshchalkov, Condat, Marconi, Giojalas, and
  Silhanek}}]{Guidobaldi_2014}
\bibinfo{author}{\bibfnamefont{A.}~\bibnamefont{Guidobaldi}},
  \bibinfo{author}{\bibfnamefont{Y.}~\bibnamefont{Jeyaram}},
  \bibinfo{author}{\bibfnamefont{I.}~\bibnamefont{Berdakin}},
  \bibinfo{author}{\bibfnamefont{V.~V.} \bibnamefont{Moshchalkov}},
  \bibinfo{author}{\bibfnamefont{C.~A.} \bibnamefont{Condat}},
  \bibinfo{author}{\bibfnamefont{V.~I.} \bibnamefont{Marconi}},
  \bibinfo{author}{\bibfnamefont{L.}~\bibnamefont{Giojalas}}, \bibnamefont{and}
  \bibinfo{author}{\bibfnamefont{A.~V.} \bibnamefont{Silhanek}},
  \bibinfo{journal}{Physics Review E,} \textbf{\bibinfo{volume}{89}},
  \bibinfo{pages}{032720} (\bibinfo{year}{2014}).

\bibitem[{\citenamefont{Xu and Jiang}(2019)}]{Xu_2019}
\bibinfo{author}{\bibfnamefont{L.}~\bibnamefont{Xu}} \bibnamefont{and}
  \bibinfo{author}{\bibfnamefont{Y.}~\bibnamefont{Jiang}},
  \bibinfo{journal}{Cells} \textbf{\bibinfo{volume}{8}}, \bibinfo{pages}{736}
  (\bibinfo{year}{2019}).

\bibitem[{vid({\natexlab{a}})}]{videoA}
\bibinfo{note}{See Supplemental Material at [URL] for video A. This is a 60
  frames per second video of a drop containing vinegar eels (T. aceiti) and
  fluorescent particles. The video shows circulation.}

\bibitem[{vid({\natexlab{b}})}]{videoB}
\bibinfo{note}{See Supplemental Material at [URL] for video B. A high speed
  1057 frames per second video of vinegar eels (T. aceiti) at high
  concentration seen under the microscope. The positions of nematode tails are
  marked with colored line segments.}

\bibitem[{\citenamefont{Allan et~al.}()\citenamefont{Allan, Caswell, Keim, and
  van~der Wel}}]{trackpy}
\bibinfo{author}{\bibfnamefont{D.}~\bibnamefont{Allan}},
  \bibinfo{author}{\bibfnamefont{T.}~\bibnamefont{Caswell}},
  \bibinfo{author}{\bibfnamefont{N.}~\bibnamefont{Keim}}, \bibnamefont{and}
  \bibinfo{author}{\bibfnamefont{C.}~\bibnamefont{van~der Wel}},
  \emph{\bibinfo{title}{Trackpy v0.3.2}},
  \urlprefix\url{{https://doi.org/10.5281/zenodo.60550}}.

\bibitem[{\citenamefont{Crocker and Grier}(1996)}]{crocker96}
\bibinfo{author}{\bibfnamefont{J.~C.} \bibnamefont{Crocker}} \bibnamefont{and}
  \bibinfo{author}{\bibfnamefont{D.~G.} \bibnamefont{Grier}},
  \bibinfo{journal}{Journal of Colloid Interface Science}
  \textbf{\bibinfo{volume}{179}}, \bibinfo{pages}{298} (\bibinfo{year}{1996}).

\bibitem[{\citenamefont{Liron and Mochon}(1976)}]{Liron_1976}
\bibinfo{author}{\bibfnamefont{N.}~\bibnamefont{Liron}} \bibnamefont{and}
  \bibinfo{author}{\bibfnamefont{S.}~\bibnamefont{Mochon}},
  \bibinfo{journal}{Journal of Engineering Mathematics}
  \textbf{\bibinfo{volume}{10}}, \bibinfo{pages}{287} (\bibinfo{year}{1976}).

\bibitem[{\citenamefont{Brenner}(1999)}]{Brenner_1999}
\bibinfo{author}{\bibfnamefont{M.~P.} \bibnamefont{Brenner}},
  \bibinfo{journal}{Physics of Fluids} \textbf{\bibinfo{volume}{11}},
  \bibinfo{pages}{754} (\bibinfo{year}{1999}).

\bibitem[{\citenamefont{Cui et~al.}(2002)\citenamefont{Cui, Diamant, and
  Lin}}]{Cui_2002}
\bibinfo{author}{\bibfnamefont{B.}~\bibnamefont{Cui}},
  \bibinfo{author}{\bibfnamefont{H.}~\bibnamefont{Diamant}}, \bibnamefont{and}
  \bibinfo{author}{\bibfnamefont{B.}~\bibnamefont{Lin}},
  \bibinfo{journal}{Physical Review Letters} \textbf{\bibinfo{volume}{89}},
  \bibinfo{pages}{188302} (\bibinfo{year}{2002}).

\bibitem[{\citenamefont{Batchelor}(2000)}]{batchelor2000introduction}
\bibinfo{author}{\bibfnamefont{G.~K.} \bibnamefont{Batchelor}},
  \emph{\bibinfo{title}{An introduction to fluid dynamics}}
  (\bibinfo{publisher}{Cambridge University Press (Cambridge, UK)},
  \bibinfo{year}{2000}).

\bibitem[{\citenamefont{Zeng et~al.}(2003)\citenamefont{Zeng, Yortsos, and
  Salin}}]{zeng2003brinkman}
\bibinfo{author}{\bibfnamefont{J.}~\bibnamefont{Zeng}},
  \bibinfo{author}{\bibfnamefont{Y.~C.} \bibnamefont{Yortsos}},
  \bibnamefont{and} \bibinfo{author}{\bibfnamefont{D.}~\bibnamefont{Salin}},
  \bibinfo{journal}{Physics of Fluids} \textbf{\bibinfo{volume}{15}},
  \bibinfo{pages}{3829} (\bibinfo{year}{2003}).

\bibitem[{\citenamefont{Boos and Thess}(1997)}]{boos1997thermocapillary}
\bibinfo{author}{\bibfnamefont{W.}~\bibnamefont{Boos}} \bibnamefont{and}
  \bibinfo{author}{\bibfnamefont{A.}~\bibnamefont{Thess}},
  \bibinfo{journal}{Journal of Fluid Mechanics} \textbf{\bibinfo{volume}{352}},
  \bibinfo{pages}{305} (\bibinfo{year}{1997}).

\bibitem[{\citenamefont{Bush}(1997)}]{bush1997anomalous}
\bibinfo{author}{\bibfnamefont{J.~W.~M.} \bibnamefont{Bush}},
  \bibinfo{journal}{Journal of Fluid Mechanics} \textbf{\bibinfo{volume}{352}},
  \bibinfo{pages}{283} (\bibinfo{year}{1997}).

\bibitem[{\citenamefont{Mathijssen et~al.}(2016)\citenamefont{Mathijssen,
  Doostmohammadi, Yeomans, and Shendruk}}]{Mathijssen_2016}
\bibinfo{author}{\bibfnamefont{A.~J. T.~M.} \bibnamefont{Mathijssen}},
  \bibinfo{author}{\bibfnamefont{A.}~\bibnamefont{Doostmohammadi}},
  \bibinfo{author}{\bibfnamefont{J.~M.} \bibnamefont{Yeomans}},
  \bibnamefont{and} \bibinfo{author}{\bibfnamefont{T.~N.}
  \bibnamefont{Shendruk}}, \bibinfo{journal}{Journal of Fluid Mechanics}
  \textbf{\bibinfo{volume}{806}}, \bibinfo{pages}{35} (\bibinfo{year}{2016}).

\bibitem[{\citenamefont{Fortune}(2022)}]{Fortune_2022}
\bibinfo{author}{\bibfnamefont{G.~T.} \bibnamefont{Fortune}}, Ph.D. thesis,
  \bibinfo{school}{University of Cambridge} (\bibinfo{year}{2022}).

\bibitem[{\citenamefont{Brinkman}(1949)}]{brinkman1949calculation}
\bibinfo{author}{\bibfnamefont{H.~C.} \bibnamefont{Brinkman}},
  \bibinfo{journal}{Flow, Turbulence and Combustion}
  \textbf{\bibinfo{volume}{1}}, \bibinfo{pages}{27} (\bibinfo{year}{1949}).

\bibitem[{\citenamefont{Purcell}(1977)}]{Purcell_1977}
\bibinfo{author}{\bibfnamefont{E.}~\bibnamefont{Purcell}},
  \bibinfo{journal}{Am. J. Phys.} \textbf{\bibinfo{volume}{45}},
  \bibinfo{pages}{3} (\bibinfo{year}{1977}).

\bibitem[{\citenamefont{Shapere and Wilczek}(1989)}]{Shapere_1989}
\bibinfo{author}{\bibfnamefont{A.}~\bibnamefont{Shapere}} \bibnamefont{and}
  \bibinfo{author}{\bibfnamefont{F.}~\bibnamefont{Wilczek}},
  \bibinfo{journal}{J. Fluid Mech.} \textbf{\bibinfo{volume}{198}},
  \bibinfo{pages}{557} (\bibinfo{year}{1989}).

\bibitem[{\citenamefont{Ehlers and Koiller}(2011)}]{Ehlers_2011}
\bibinfo{author}{\bibfnamefont{K.~M.} \bibnamefont{Ehlers}} \bibnamefont{and}
  \bibinfo{author}{\bibfnamefont{J.}~\bibnamefont{Koiller}},
  \bibinfo{journal}{Regular and Chaotic Dynamics}
  \textbf{\bibinfo{volume}{16}}, \bibinfo{pages}{1} (\bibinfo{year}{2011}).

\bibitem[{\citenamefont{Sznitman et~al.}(2010)\citenamefont{Sznitman, Shen,
  Sznitman, and Arratia}}]{Sznitman_2010}
\bibinfo{author}{\bibfnamefont{J.}~\bibnamefont{Sznitman}},
  \bibinfo{author}{\bibfnamefont{X.}~\bibnamefont{Shen}},
  \bibinfo{author}{\bibfnamefont{R.}~\bibnamefont{Sznitman}}, \bibnamefont{and}
  \bibinfo{author}{\bibfnamefont{P.~E.} \bibnamefont{Arratia}},
  \bibinfo{journal}{Physics of Fluids} \textbf{\bibinfo{volume}{22}},
  \bibinfo{pages}{121901} (\bibinfo{year}{2010}).

\bibitem[{\citenamefont{Majmudar et~al.}(2012)\citenamefont{Majmudar, Keaveny,
  Zhang, and Shelley}}]{majmudar2012experiments}
\bibinfo{author}{\bibfnamefont{T.}~\bibnamefont{Majmudar}},
  \bibinfo{author}{\bibfnamefont{E.~E.} \bibnamefont{Keaveny}},
  \bibinfo{author}{\bibfnamefont{J.}~\bibnamefont{Zhang}}, \bibnamefont{and}
  \bibinfo{author}{\bibfnamefont{M.~J.} \bibnamefont{Shelley}},
  \bibinfo{journal}{Journal of the Royal Society Interface}
  \textbf{\bibinfo{volume}{9}}, \bibinfo{pages}{1809} (\bibinfo{year}{2012}).

\bibitem[{\citenamefont{Keller and Rubinow}(1976)}]{KR1976}
\bibinfo{author}{\bibfnamefont{J.~B.} \bibnamefont{Keller}} \bibnamefont{and}
  \bibinfo{author}{\bibfnamefont{S.~I.} \bibnamefont{Rubinow}},
  \bibinfo{journal}{Journal of Fluid Mechanics} \textbf{\bibinfo{volume}{75}},
  \bibinfo{pages}{705} (\bibinfo{year}{1976}).

\bibitem[{\citenamefont{Tornberg and Shelley}(2004)}]{tornberg2004simulating}
\bibinfo{author}{\bibfnamefont{A.-K.} \bibnamefont{Tornberg}} \bibnamefont{and}
  \bibinfo{author}{\bibfnamefont{M.~J.} \bibnamefont{Shelley}},
  \bibinfo{journal}{Journal of Computational Physics}
  \textbf{\bibinfo{volume}{196}}, \bibinfo{pages}{8} (\bibinfo{year}{2004}).

\bibitem[{\citenamefont{Backholm et~al.}(2013)\citenamefont{Backholm, Ryu, and
  Dalnoki-Veress}}]{backholm2013viscoelastic}
\bibinfo{author}{\bibfnamefont{M.}~\bibnamefont{Backholm}},
  \bibinfo{author}{\bibfnamefont{W.~S.} \bibnamefont{Ryu}}, \bibnamefont{and}
  \bibinfo{author}{\bibfnamefont{K.}~\bibnamefont{Dalnoki-Veress}},
  \bibinfo{journal}{Proceedings of the National Academy of Sciences,}
  \textbf{\bibinfo{volume}{110}}, \bibinfo{pages}{4528} (\bibinfo{year}{2013}).

\bibitem[{\citenamefont{Chakrabarti and
  Saintillan}(2019)}]{chakrabarti2019spontaneous}
\bibinfo{author}{\bibfnamefont{B.}~\bibnamefont{Chakrabarti}} \bibnamefont{and}
  \bibinfo{author}{\bibfnamefont{D.}~\bibnamefont{Saintillan}},
  \bibinfo{journal}{Physical Review Fluids} \textbf{\bibinfo{volume}{4}},
  \bibinfo{pages}{043102} (\bibinfo{year}{2019}).

\bibitem[{\citenamefont{Canio et~al.}(2017)\citenamefont{Canio, Lauga, and
  Goldstein}}]{Canio_2017}
\bibinfo{author}{\bibfnamefont{G.~D.} \bibnamefont{Canio}},
  \bibinfo{author}{\bibfnamefont{E.}~\bibnamefont{Lauga}}, \bibnamefont{and}
  \bibinfo{author}{\bibfnamefont{R.~E.} \bibnamefont{Goldstein}},
  \bibinfo{journal}{J. R. Soc. Interface} \textbf{\bibinfo{volume}{14}},
  \bibinfo{pages}{20170491} (\bibinfo{year}{2017}).

\bibitem[{\citenamefont{Wen et~al.}(2012)\citenamefont{Wen, Po, Hulme, Chen,
  Liu, Kwok, Gershow, Leifer, Butler, Fang-Yen et~al.}}]{Wen_2012}
\bibinfo{author}{\bibfnamefont{Q.}~\bibnamefont{Wen}},
  \bibinfo{author}{\bibfnamefont{M.~D.} \bibnamefont{Po}},
  \bibinfo{author}{\bibfnamefont{E.}~\bibnamefont{Hulme}},
  \bibinfo{author}{\bibfnamefont{S.}~\bibnamefont{Chen}},
  \bibinfo{author}{\bibfnamefont{X.}~\bibnamefont{Liu}},
  \bibinfo{author}{\bibfnamefont{S.~W.} \bibnamefont{Kwok}},
  \bibinfo{author}{\bibfnamefont{M.}~\bibnamefont{Gershow}},
  \bibinfo{author}{\bibfnamefont{A.~M.} \bibnamefont{Leifer}},
  \bibinfo{author}{\bibfnamefont{V.}~\bibnamefont{Butler}},
  \bibinfo{author}{\bibfnamefont{C.}~\bibnamefont{Fang-Yen}},
  \bibnamefont{et~al.}, \bibinfo{journal}{Neuron}
  \textbf{\bibinfo{volume}{76}}, \bibinfo{pages}{750} (\bibinfo{year}{2012}).

\bibitem[{\citenamefont{Hao et~al.}(2022)\citenamefont{Hao, Zhou, and
  Gravish}}]{Hao_2022}
\bibinfo{author}{\bibfnamefont{Z.}~\bibnamefont{Hao}},
  \bibinfo{author}{\bibfnamefont{W.}~\bibnamefont{Zhou}}, \bibnamefont{and}
  \bibinfo{author}{\bibfnamefont{N.}~\bibnamefont{Gravish}},
  \bibinfo{journal}{Advanced Robotics} \textbf{\bibinfo{volume}{36}},
  \bibinfo{pages}{1} (\bibinfo{year}{2022}).

\bibitem[{\citenamefont{Taylor}(1951)}]{Taylor_1951}
\bibinfo{author}{\bibfnamefont{G.~I.} \bibnamefont{Taylor}},
  \bibinfo{journal}{Proceedings of the Royal Society of London. Series A.
  Mathematical and Physical Sciences} \textbf{\bibinfo{volume}{209}},
  \bibinfo{pages}{447} (\bibinfo{year}{1951}),
  \urlprefix\url{https://doi.org/10.1098%2Frspa.1951.0218}.

\bibitem[{\citenamefont{Brennan}(1974)}]{Brennen_1974}
\bibinfo{author}{\bibfnamefont{C.}~\bibnamefont{Brennan}},
  \bibinfo{journal}{Journal of Fluid Mechanics} \textbf{\bibinfo{volume}{65}},
  \bibinfo{pages}{799} (\bibinfo{year}{1974}).

\bibitem[{\citenamefont{Koiller et~al.}(1996)\citenamefont{Koiller, Ehlers, and
  Montgomery}}]{Koiller_1996}
\bibinfo{author}{\bibfnamefont{J.}~\bibnamefont{Koiller}},
  \bibinfo{author}{\bibfnamefont{K.}~\bibnamefont{Ehlers}}, \bibnamefont{and}
  \bibinfo{author}{\bibfnamefont{R.}~\bibnamefont{Montgomery}},
  \bibinfo{journal}{J. Nonlinear Sci.} \textbf{\bibinfo{volume}{6}},
  \bibinfo{pages}{507} (\bibinfo{year}{1996}).

\bibitem[{\citenamefont{Chelakkot et~al.}(2021)\citenamefont{Chelakkot, Hagan,
  and Gopinath}}]{chelakkot2021synchronized}
\bibinfo{author}{\bibfnamefont{R.}~\bibnamefont{Chelakkot}},
  \bibinfo{author}{\bibfnamefont{M.~F.} \bibnamefont{Hagan}}, \bibnamefont{and}
  \bibinfo{author}{\bibfnamefont{A.}~\bibnamefont{Gopinath}},
  \bibinfo{journal}{Soft matter} \textbf{\bibinfo{volume}{17}},
  \bibinfo{pages}{1091} (\bibinfo{year}{2021}).

\end{thebibliography}

\vfill\eject

\

\appendix

\section{Computing viscous dissipation}
\label{sec:power}



The viscous dissipation rate in an incompressible flow depends on 
the traceless part of the velocity gradient 
\begin{equation}
\sigma_{ij} = \frac{1}{2} (\partial_{x_j} u_{i} 
+ \partial_{x_i} u_{j}) . 
\end{equation}
The viscous dissipation rate per unit volume 
\begin{align}
\dot \epsilon &= \mu\ {\rm tr} \sigma^2 = \mu \sum_{ij} (\sigma_{ij})^2 
\end{align}
where the dynamic viscosity $\mu = \rho \nu$, 
and $\nu$ is the kinematic viscosity. 

We integrate over $z$ to find
the viscous dissipation rate per unit area.

It is useful to compute dimensionless constants that depend on the assumed form for the velocity field as a function of depth 
\begin{align}
\bar f &= \frac{1}{h} \int_0^h f(z) dz\\
c_0 &=\frac{1}{h} \int_0^h f(z)^2 dz \\
c_1 &= h \int_0^h f'(z)^2 dz. \label{eqn:c0c1}
\end{align}

The full three-dimensional velocity field $(U,V,W)$
is related to the vertically averaged two-dimensional velocity field $(u,v)$ with 
\begin{align}
U(x,y,z,t) &  =\frac{1}{\bar f} u(x,y,t) f(z)\nonumber \\
V(x,y,z,t) & =\frac{1}{\bar f} v(x,y,t) f(z) \nonumber \\
W(x,y,z,t) & = 0 .
\end{align}
From $U,V,W$,
we compute the different components of the velocity gradient tensor in terms of the stream function 
\begin{align}
\sigma_{xx} &= \partial_x U =  \frac{f(z) }{\bar f}  \partial_x u 
 = \frac{ f(z)}{\bar f}   \partial_{xy} \psi\nonumber \\
\sigma_{yy} & = \partial_x V =  \frac{ f(z) }{\bar f}  f(z) \partial_y v
 = -\frac{ f(z)}{\bar f}   \partial_{xy} \psi \nonumber \\
\sigma_{zz} & = \partial_z W = 0 \nonumber \\
\sigma_{xz} & = \frac{1}{2} ( \partial_x W + \partial_z U) = \frac{ f'(z) }{2 \bar f} u = 
= \frac{ f'(z) }{2 \bar f} \partial_y \psi \nonumber \\
\sigma_{yz} & = \frac{1}{2} ( \partial_y W + \partial_z V) = \frac{ f'(z) }{2 \bar f} v 
= -\frac{ f'(z) }{2 \bar f} \partial_x \psi\nonumber \\
\sigma_{xy} & = \frac{1}{2} \partial_x V + \partial_y U) = \frac{ f(z)}{2 \bar f}
( \partial_x v + \partial_y u)\nonumber  \\
& = \frac{ f(z)}{2 \bar f}
(- \partial_{xx} \psi + \partial_{yy} \psi) .
\end{align}

We integrate the squares of the components of the velocity gradient over depth;
\begin{align}
\int_0^h dz \  \sigma_{xx}^2 &= 
 \frac{\int_0^h  f(z)^2 dz}{(\bar f)^2}   (\partial_{xy} \psi)^2 \nonumber \\
 & = \frac{h c_0}{(\bar f)^2} (\partial_{xy} \psi)^2  \nonumber \\
\int_0^h dz \  \sigma_{yy}^2 &=
   \frac{h c_0}{(\bar f)^2} (\partial_{xy} \psi)^2   \nonumber \\
\int_0^h dz \  \sigma_{zz}^2 &= 0 \nonumber \\
\int_0^h dz \  \sigma_{xy}^2 &= \frac{\int_0^h  f(z)^2 dz}{(\bar f)^2} \frac{1}{4} ( \partial_{yy}\psi - \partial_{xx} \psi)^2 \nonumber \\
& = \frac{h c_0}{(\bar f)^2} \frac{1}{4}( \partial_{yy}\psi - \partial_{xx} \psi)^2 \nonumber \\
\int_0^h dz \  \sigma_{xz}^2 &=
\frac{\int_0^h  f'(z)^2 dz}{(\bar f)^2} \frac{1}{4} (\partial_y \psi)^2 \nonumber \\
& = \frac{ c_1}{ h (\bar f)^2}\frac{1}{4} 
(\partial_y \psi)^2 \nonumber \\
\int_0^h dz \  \sigma_{yz}^2 
& = \frac{ c_1}{h (\bar f)^2}\frac{1}{4} 
(\partial_x \psi)^2.
\end{align}

In terms of the stream function, 
the viscous dissipation rate per unit area
is 
\begin{align}
{\dot e}_A(x,y)\! =& \!
    \int_0^h dz\ {\dot \epsilon }(x,y,z) \nonumber \\
    =& \mu h \Bigg[\frac{c_0}{(\bar f)^2} \left( 2 (\partial_{xy} \psi)^2
    + \frac{1}{2} (\partial_{yy} \psi - \partial_{xx} \psi)^2\right)\nonumber \\
    & +  \frac{c_1}{2 h^2 (\bar f)^2 } \left( (\partial_y \psi)^2
    + (\partial_x \psi)^2 \right) \Bigg]. \label{eqn:dotea}
\end{align}
Using $f(z)$ from Eqn.~\ref{eqn:fz}, 
corresponding to a fixed base and free
surface, and with Eqns.~\ref{eqn:c0c1}, we find  
${\bar f} = 2/3$, $c_0 =$ 8/15, $c_1 =$ 4/3, 
$c_0/(\bar f)^2 = 1.2$, and $c_1/(\bar f)^2 = 3$ which are used in Eqn.\ref{eqn:dotea}.

To estimate the viscous dissipation rate, 
we integrate the dissipation rate per unit area
with Eqn.~\ref{eqn:dotea}, over an area with $x$ ranging from 0 to $\lambda_{MW}$ and $y$ ranging from the boundary to $y=y_m +  \lambda_{MW}$ where $y_m$ (defined in Eqn.~\ref{eqn:b}) is approximately the mean $y$ value for the boundary.

\end{document}